\documentclass[3p]{elsarticle}

\usepackage{hyperref}
\usepackage{amsmath}
\usepackage{mathtools}
\usepackage{subfig}
\usepackage{floatrow}  
\usepackage{csquotes}
\usepackage{color,soul}
\usepackage{titlesec}
\usepackage[shortlabels]{enumitem}
\usepackage{multirow}
\usepackage{amsfonts}
\usepackage{amssymb}
\usepackage{tikz}
\usepackage{tabularx}
\usepackage{adjustbox}
\usepackage{url}
\usepackage{makecell}
\usepackage{bm}
\usepackage[normalem]{ulem}
\usepackage{floatrow}
\usepackage{todonotes}

\journal{ISPRS Journal of Photogrammetry and Remote Sensing}

\begin{document}
\sloppy
\begin{frontmatter}

\title{Large-scale Building Height Retrieval from Single SAR Imagery based on Bounding Box Regression Networks}

\author[1a]{Yao Sun}
\ead{yao.sun@dlr.de}

\author[1a,2a]{Lichao Mou}
\ead{lichao.mou@dlr.de}

\author[2a]{Yuanyuan Wang}
\ead{yuanyuan.wang@dlr.de} 

\author[2a]{Sina Montazeri}
%\ead{sina.montazeri1988@gmail.com}

\author[1a,2a]{Xiao Xiang Zhu\corref{mycorrespondingauthor}}
\cortext[mycorrespondingauthor]{Corresponding author}
\ead{xiaoxiang.zhu@dlr.de}

\address[1a]{Department of Aerospace and Geodesy, Data Science in Earth Observation, Technical University of Munich, Arcisstraße 21, 80333 Munich, Germany}
\address[2a]{Remote Sensing Technology Institute, German Aerospace Center (DLR), Münchener Straße 20, 82234 Weßling, Germany}

\begin{abstract}
Building height retrieval from synthetic aperture radar (SAR) imagery is of great importance for urban applications, yet highly challenging owing to the complexity of SAR data. 
This paper addresses the issue of building height retrieval in large-scale urban areas from a single TerraSAR-X spotlight or stripmap image. 
Based on the radar viewing geometry, we propose that this problem can be formulated as a bounding box regression problem and therefore allows for integrating height data from multiple data sources in generating ground truth on a larger scale. 
We introduce building footprints from {Geographic information system (GIS)} data as complementary information and propose a bounding box regression network that exploits the location relationship between a building's footprint and its bounding box, allowing for fast computation. This is important for large-scale applications. The method is validated on four urban data sets using TerraSAR-X images in both high-resolution spotlight and stripmap modes. 
Experimental results show that 
the proposed network can reduce the computation cost significantly while keeping the height accuracy of individual buildings compared to a Faster R-CNN based method. 
Moreover, we investigate the impact of inaccurate GIS data on our proposed network, and this study shows that the bounding box regression network is robust against positioning errors in GIS data. 
The proposed method has great potential to be applied to regional or even global scales. 
\end{abstract}


\end{frontmatter}

%\linenumbers

%%%%%%% MAIN %%%%%%%%%%%
\section{Introduction}\label{sec:intro}

  \begin{figure*}[!]
     \centering
     \subfloat[SAR image]{\includegraphics[ height=0.34\linewidth,  width=0.24\linewidth]{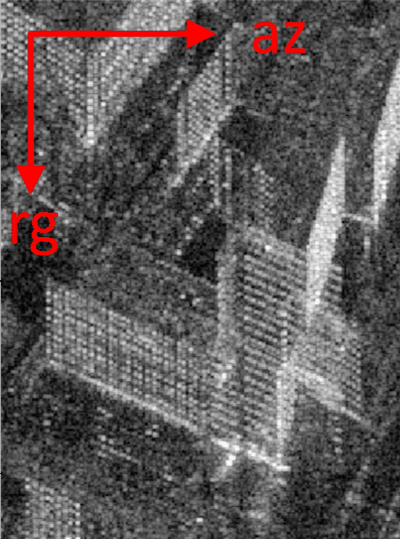}}\hfill
     \subfloat[Building footprints]{\includegraphics[ height=0.34\linewidth, width=0.24\linewidth]{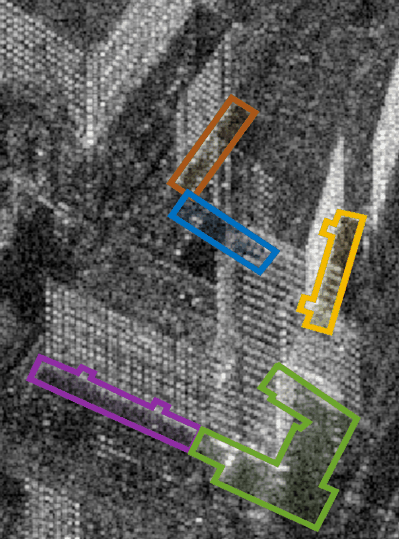}}\hfill
     \subfloat[Building bounding boxes]{\includegraphics[ height=0.34\linewidth, width=0.24\linewidth]{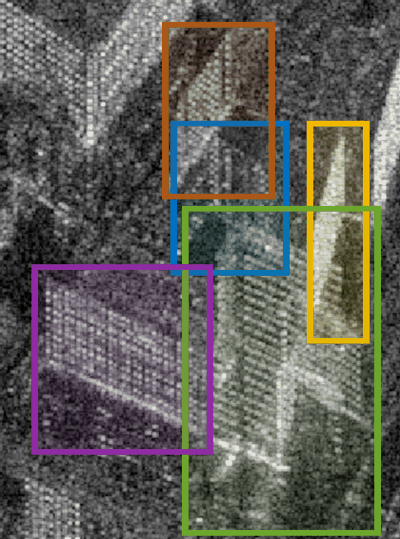}}\hfill
     \subfloat[LoD1 building models]{\includegraphics[ height=0.34\linewidth, width=0.24\linewidth]{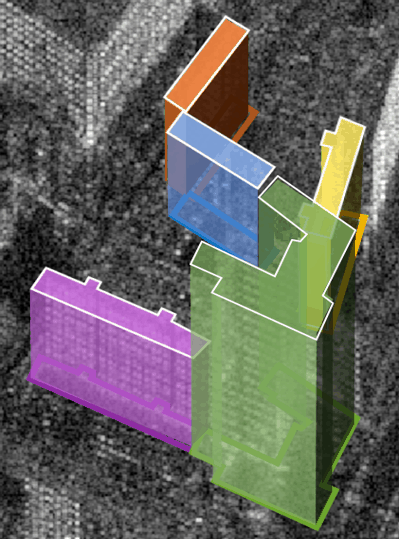}}
     \caption{Illustration of the input and output of our method. 
     (a) and (b) are the input data: a SAR image and building footprints in the SAR image. (c) shows the predicted bounding boxes of these buildings. Building heights are then computed from the bounding boxes and building footprints, and levels-of-detail (LoD) 1 models are reconstructed, as shown in (d). 
     rg and az denote the slant range direction and azimuth direction, respectively.  }
     \label{fig:f1}
 \end{figure*}

Three-dimensional (3-D) building models are widely used in public and commercial sectors for environmental researches and location-based services.  
For the past three decades, 3-D building reconstruction has been a hot topic in remote sensing~\cite{rottensteiner2012isprs}; however, there is limited information on the third dimension, i.e., building height, on a regional or global scale. 
Studies on building height retrieval primarily employ high-resolution optical images and airborne LiDAR data~\cite{brenner2005building}. Optical data acquisition requires the weather to be cloud-free, and airborne or terrestrial data are too expensive to collect globally. 

Synthetic Aperture Radar (SAR) imagery, on the other hand, is capable of providing data regardless of time or weather conditions. Such data are of great interest to applications of disaster responses~\cite{brunner2010Earthquake, wang2012Postearthquake} and to studies concerning regions frequently covered by clouds~\cite{huang2015Cloud}. 
Since the launch of TerraSAR-X in 2007, modern SAR satellites, e.g., TerraSAR-X, TanDEM-X, and CosmoSky-Med, have been providing meter or even sub-meter resolution images, making it possible to extract and reconstruct man-made objects from spaceborne SAR data. 
In addition, complete global coverages of TerraSAR-X and Tandem-X stripmap mode data have been acquired since 2012, providing great potential as a data source for global building reconstruction~\cite{8437985}. 

The study of building analysis from SAR imagery dates back to 1969, 
that Laprade and Leonardo derive the elevation of a few buildings from shadows and layovers using simulated radar images~\cite{laprade1969elevations}.  
Since then, various studies have been conducted on this topic~\cite{franceschetti2002canonical, tupin2003detection, guida2010height, brunner2010Buildinga, sportouche2011Extraction, wenliu2013Building}.
However, building interpretation from SAR data is highly challenging. 
Due to the side-looking geometry and one-band radar sensors, urban structures are clearly visible in SAR images but are difficult to distinguish from each other. 
Several works~\cite{guida2010height, sportouche2011Extraction, brunner2010Building} develop tailored algorithms for building analyses in complex urban environments, but these methods are limited to be applied for large-scale areas. 
Recently, deep neural networks are applied for individual building segmentation in SAR images~\cite{sun2020cgnet}. Building heights are subsequently computed based on the radar viewing geometry. 
However, for annotating building areas, this method requires an accurate digital elevation model (DEM), which is unavailable in most areas and thus restricts this method from being generalized to other regions.

In this work, we are interested in the height estimation of individual buildings on a large scale, using single SAR images. 
We develop a method that takes SAR images and building footprints as input and retrieves building heights by predicting bounding boxes of buildings (cf. Figure~\ref{fig:f1}). 
Next, we briefly explain the challenges involved in this task and review related work. 

\subsection{Challenges} 

\begin{figure*}[!]
    \centering    
    \subfloat[high-rise building]{\includegraphics[trim=0.8cm 0cm 0.8cm 0cm, clip=true, width=0.34\linewidth]{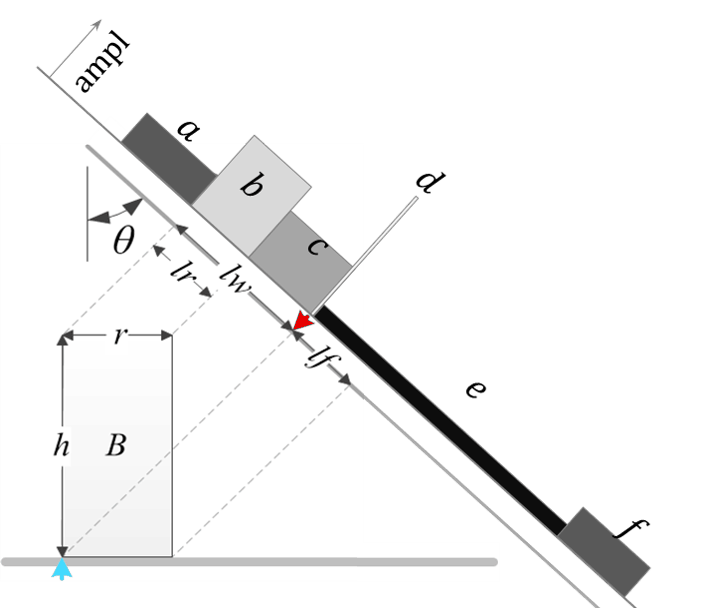}}\hspace{2cm}
    \subfloat[low-rise building]{\includegraphics[trim=0.7cm 0cm 1.2cm 0cm, clip=true,width=0.33\linewidth]{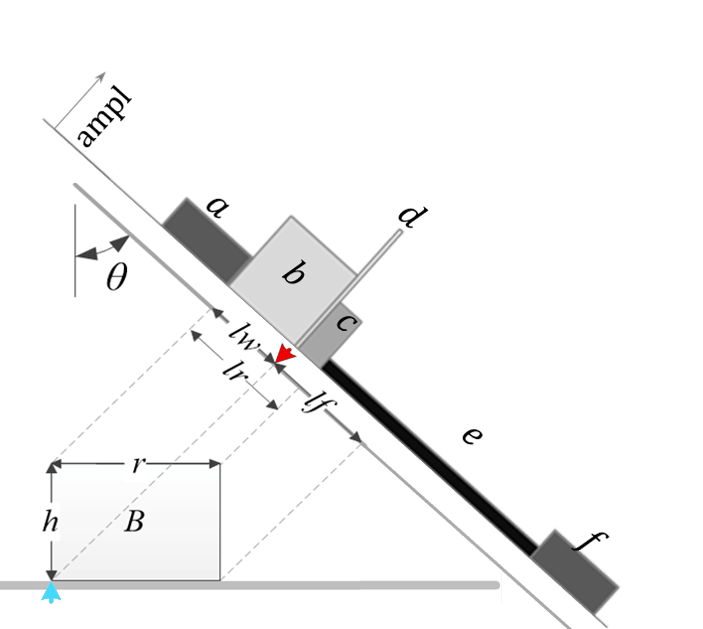}}
    \caption{Illustration of the amplitude profile (ampl) of two flat-roof buildings in a slant-range SAR image. 
    $\theta$ is the incidence angle. $h$ is the building height. 
    \textit{lw}, \textit{lr}, and \textit{lf} denote the areas of wall, roof, and footprint in the slant-range SAR image, respectively. 
    The gray shades and heights of regions \textit{a-f} indicate the expected magnitude values of intensity on the SAR image. 
    The blue arrow marks the bottom of the sensor-facing wall and the red arrow points at the double bounce line on the SAR image.}
    \label{fig:b1_b2}
\end{figure*}

Because of the side-looking imaging geometry and complex backscattering mechanism, SAR image interpretation is a generally difficult task.  
For interpreting individual buildings in urban SAR images, the challenges are mainly two-fold: 

For \textit{an isolated building} in SAR images, the main challenge is to recognize its components, i.e., the roof, walls, and footprint. 
Figure~\ref{fig:b1_b2} illustrates the amplitude profile of two flat-roof buildings in a slant-range SAR image. 
As can be seen, the wall area \textit{lw} and the roof area \textit{lr} in SAR images are always mixed and difficult to differentiate: 
\textit{lw} covers \textit{lr} when the building height \textit{h} is large (cf. Figure~\ref{fig:b1_b2} (a)), 
  and it is covered by \textit{lr} when $h$ is small (cf. Figure~\ref{fig:b1_b2} (b)). 
  In addition, for low-rise buildings, 
  the roof area \textit{lr} partially overlaps the footprint area \textit{lf} (cf. Figure~\ref{fig:b1_b2} (b)), and therefore the near-range side of \textit{lf} might be ambiguous. 
  Moreover, the far-range side of the footprint area \textit{lf} is unknown, as it connects the shadow area that also appears dark in SAR images.

For \textit{multiple adjacent or nearby buildings}, a more crucial issue is to identify them correctly. 
Since the intensity values in SAR images are closely related to material types and structural shapes of objects, consecutive buildings in the physical world are difficult to separate in a SAR image unless obvious material or structure changes exist at building boundaries.  
In addition, even if buildings in the real world are not neighboring, they may overlap each other in SAR images, which significantly increases the difficulty of image interpretation. 
For example, Figure~\ref{fig:f1} (a) shows a typical urban region in a TerraSAR-X spotlight image containing several buildings whose footprints and bounding boxes are plotted in Figure~\ref{fig:f1} (b) and (c), respectively. 
By only looking at the SAR image, it is unlikely to tell the numbers of buildings or distinguish connected buildings, for instance, the purple building and the green building. 
Besides, it is noticeable that the green building overlaps the yellow and the blue ones in the SAR image, although their footprints are not connected.

\subsection{Related work}

In literature, researchers have investigated building height retrieval from 
a single SAR image~\cite{laprade1969elevations, quartulli2004Stochastic, brunner2010Buildinga, zhao2013Building}, 
InSAR data~\cite{thiele_extraction_2013, dubois2016Building}, 
and multi-aspect SAR or InSAR data~\cite{leberl1989radargrammetric, soergel_stereo_2009, xu2007Automatic} or even circular SAR~\cite{oriot2008circular, 6198884} to overcome the drawback of occlusions originated from the side-looking geometry. 
In addition to SAR data, auxiliary data, e.g., building outlines extracted from optical images~\cite{wegner2009Building, sportouche2011Extraction} and building footprints obtained from GIS data~\cite{thiele2010Combining, wenliu2013Building, yaosun2017Building}, are introduced for providing exact locations and geometric shapes of buildings in the real world. 

To retrieve building heights from SAR data, 
most researchers employ data-driven approaches that first extract salient features such as double bounce lines, layovers, shadows, and InSAR phases and then deduce building parameters. 
Some researchers first detect bright-line segments and regularly spaced point-like features and subsequently group them into building regions~\cite{tupin2003detection, xu2007Automatic, michaelsen2006Perceptual, soergel2009Stereo, ferro2013Automatic}. 
Alternatively, several studies directly extract building regions of layovers and shadows using segmentation algorithms, 
such as the mean shift algorithm and conditional random field (CRF)~\cite{he2008building}, 
marker controlled watershed algorithm~\cite{zhao2013Building},  
thresholding and morphological operations~\cite{cao2014Detecting}. 
Besides data-driven approaches, model-based methods are conducted. 
Detailed modeling of the geometrical and radiometric properties of isolated buildings is performed in ~\cite{quartulli2004Stochastic, guida2008ModelBased, guida2010height}, 
from which building shapes in SAR data are estimated. 
Such techniques require extensive prior knowledge about objects, such as materials, roughness, humidity, and orientation with respect to the SAR sensor, which is generally unknown. 
Another model-based approach is developed in a simulating and matching fashion~\cite{jahangir2007extracting, sportouche2009Buildinga, brunner2010Building, thiele2012gis, wang2015building}.  
This approach comprises a simulation step in which a SAR image or InSAR phase is simulated using a building hypothesis and a matching step in which the simulated data are matched with real data. 
The process is conducted iteratively until the optimal building parameters are achieved.

    Although a considerable amount of research has been carried out on building height retrieval using SAR data, few studies have investigated this problem on a large scale. 
    Most methods target buildings with specific shapes, 
    e.g., rectangular-~\cite{simonetto2005Rectangular, wang2008Building,liu2017BottomUp} or L-shaped footprints~\cite{zhang2011Building, zhao2013Building}, flat~\cite{wegner2014Combining} or gable roofs~\cite{thiele2010Analysis, chen2017Automatic}, and different heights~\cite{chen2017Automatic,guo2014HighRise,liu2015Height,tang2016Highrise}.
    The majority of studies investigate simple scenarios where a minimal distance between buildings is required to ensure scattering effects of different buildings do not interfere with each other~\cite{guida_height_2010, brunner2010Buildinga, sportouche2011Extraction, shanshanchen2015Automatic}. 
    Moreover, the performance of the presented methods is typically presented for a small set of test data, usually comprising only one or a few buildings. 
    The generalisability of much published research on this issue is therefore problematic.

In recent years, deep neural networks have been becoming increasingly popular and triggered breakthroughs in many fields, including a wide range of remote sensing applications~\cite{zhu2021deep,zhang2016deep,mou2019relation,kussul2017deep,li2019r, cheng2018deep,mou2018im2height,audebert2018beyond,chen2021maskH}.   
In contrast to classical approaches that require expert domain knowledge and hand-crafted features, deep networks rely on a large amount of raw data to learn effective feature representations in an end-to-end fashion. 
But the major problem preventing applying deep networks to urban SAR analysis tasks is the lack of annotation data. 
To address this issue, 
Shahzad \textit{\textit{et al.}}~\cite{shahzad2019Buildings} introduce a SAR tomography (TomoSAR) point cloud to acquire building areas in a SAR image and take them as ground truth annotations to train a segmentation network to extract building areas. 
In~\cite{sun2019Largescale}, Sun \textit{et al.} generate building areas in a SAR image using a DEM instead of a TomoSAR point cloud, as the latter is rare. 
However, both two works do not annotate individual buildings. 
Shermeyer \textit{et al.}~\cite{shermeyer2020spacenet} present a multi-sensor all weather mapping (MSAW) dataset containing airborne SAR images, high-resolution optical images, and building footprint annotations.  
However, building footprints, instead of building heights, are the learning target in this work. 
In~\cite{sun2020cgnet}, 
Sun \textit{et al.} annotate individual buildings in a TerraSAR-X spotlight image employing a highly accurate DEM and propose a segmentation network for predicting building areas in the SAR image. The segmentation results are then applied to reconstruct building heights. 
This work has segmented individual buildings from a single SAR image on a large scale for the first time. 
However, pixel-wise labels are expensive. The data set generation approach requires accurate DEMs. 
The unavailability of accurate DEMs restricts this method from being generalized to larger areas.

\subsection{Contributions}

This work aims to retrieve building heights using a single SAR image on a large scale. 
To overcome the problem and improve transferability, 
this work proposes to generate annotation data with building heights that can be acquired from multiple sources. %,  
The task of building height estimation is formulated as a bounding box regression problem, i.e., a task to regress the center coordinate and the size of the bounding box for each building. 

The main contributions of this paper are three-fold:
\begin{enumerate}
 \item [a.]
 We propose a workflow for building height retrieval in single SAR images with GIS data. 
 To our best knowledge, this is the first time that deep networks are employed in the problem of building height retrieval in large areas from TerraSAR-X images in both high-resolution spotlight and stripmap modes. 

 \item [b.]
 We formulate the problem of building height retrieval as a bounding box regression problem and propose a bounding box regression network that is very efficient owing to the tailored use of building footprints. The fast computation speed is significant for large-scale applications. 

 \item [c.]
 We propose a ground truth generation approach to produce building bounding boxes. 
 This approach is able to integrate multiple sources of building heights, thus providing large potential in analyzing complex urban regions. 
 
\end{enumerate}

 The remainder of this paper proceeds as follows. 
 Section~\ref{sec:method} formulates the problem and delineates the proposed method.  
 Section~\ref{sec:3} is concerned with the dataset generation approach to tackle the problem of dataset scarcity.  
 The experiments and results are presented and analyzed in Section~\ref{sec:test}. 
 In Section~\ref{sec:dis}, we discuss several practical problems related to applying our method to large-scale building height retrieval. 
 Finally, Section~\ref{sec:conclude} concludes this paper.

\section{Methodology}\label{sec:method}

\subsection{Problem formulation} 

We consider LoD1 building models, i.e., prismatic models with flat roof structures \cite{kolbe2005citygml}. 
Due to the radar viewing geometry, scatterers on a vertical line in the geographic coordinate system always have the same azimuth coordinate in a SAR image, i.e., this vertical line in the SAR image parallels the range direction. 
Therefore each vertical building wall in a SAR image has one pair of opposite sides paralleling the range direction. 
This can be observed in Figure~\ref{fig:f1}. 
Hence, the extent of a building in a SAR image is bounded by two vertical lines from building walls in the azimuth direction and the region of the layover and footprint in the range direction. 
In this work, we exploit this geometric relationship to retrieve building heights. 

\begin{figure*}[h!]
            \centering
            \includegraphics[width=\linewidth]{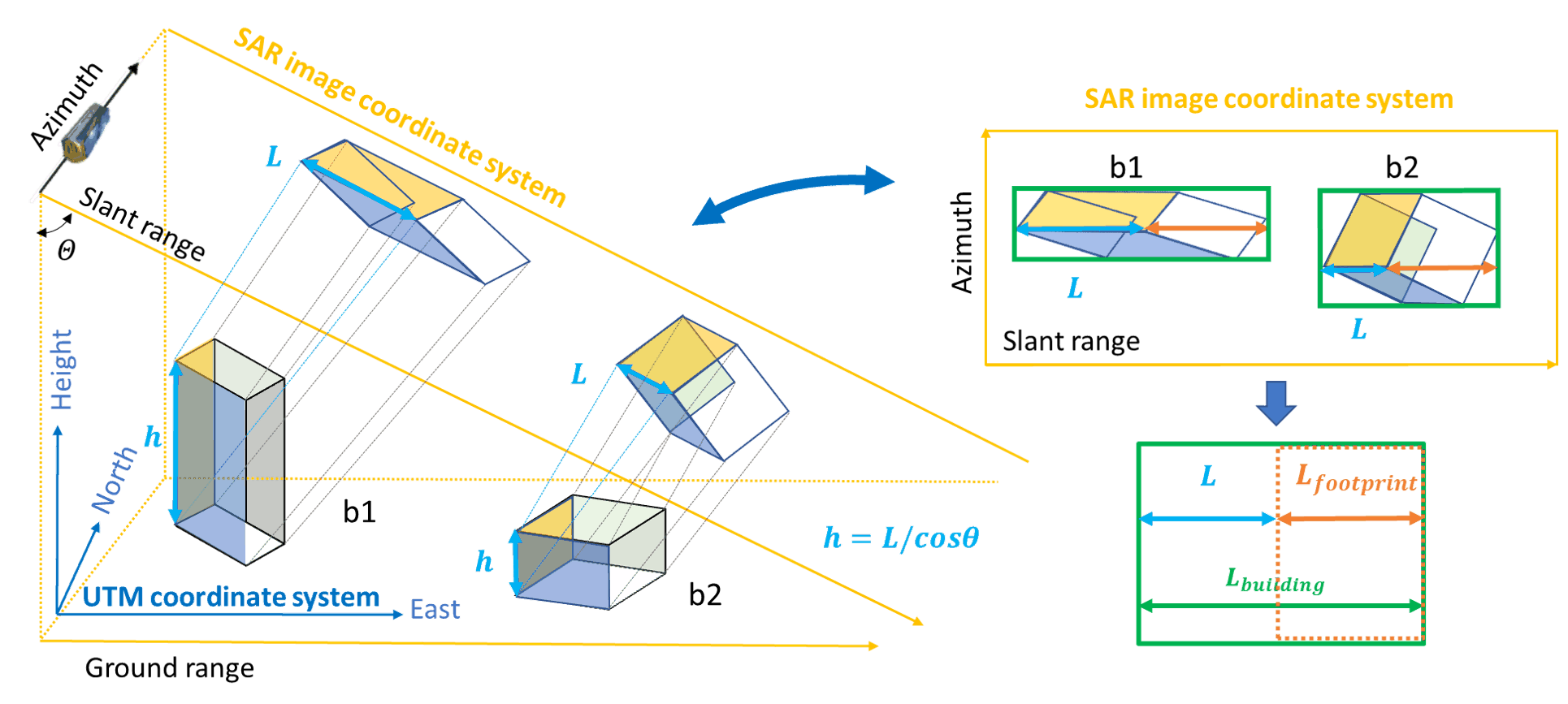}
            \caption{Illustration of bounding boxes of two buildings in a slant-range SAR image. 
            On the left, two buildings (b1 and b2) in the UTM coordinate system are imaged in a SAR image plane. %\YW{$>>$Please revise the caption accordingly$<<$}
            On the right, bounding boxes of b1 and b2 are shown in the SAR image coordinate system. 
            }
            \label{fig:b12_geo}
        \end{figure*}

Figure~\ref{fig:b12_geo} illustrates this geometric relationship by two buildings in the Universal Transverse Mercator (UTM) and the SAR image coordinate systems. On the left {of the figure}, b1 and b2 are two buildings in the UTM coordinate system and are imaged on a SAR image plane. As can be seen, sensor-visible walls (yellow and blue) are projected into the SAR image as parallelogram shapes, and vertical sides of the walls parallel the slant-range direction. The building height $h$ is directly related to the layover length $L$: 
    
    \begin{equation}
        h = L / cos\theta, 
    \end{equation}
where $\theta$ is the incidence angle.

On the right of Figure~\ref{fig:b12_geo}, b1 and b2 and their bounding boxes (green) are shown in the SAR image coordinate system. As can be seen, for both b1 and b2, the layover length $L$ is the width difference between the building bounding box and the footprint bounding box: 

    \begin{equation}
       L = L_{building} - L_{footprint}.  
    \end{equation}

Therefore, for a building in a SAR image, its height {can be} obtained once {its footprint is known and }its bounding box is detected. Based on the geometry relationships, we formulate the problem of building height retrieval from SAR images as a bounding box regression problem. 
I.e., given a SAR image and a building's footprint, find the bounding box of the building, and then derive the building height from it.  

\subsection{Footprint-guided bounding box regression}

\begin{figure*}[h!]
    \centering
    \includegraphics[width = \textwidth]{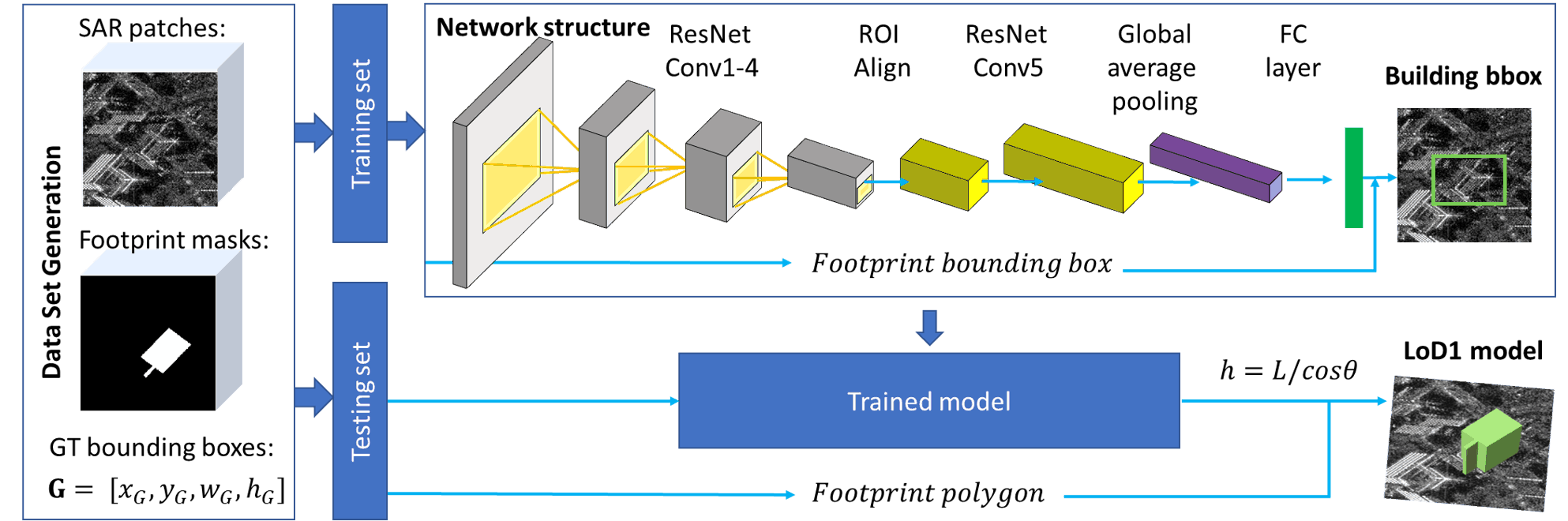}\\
    \caption{
    {General workflow of the proposed method. 
    After data set generation, our network concatenates a SAR image and a building footprint mask from the training set as input and predicts a correction for the footprint bounding box with respect to the building bounding box. 
    Then the trained model is evaluated using the testing set. 
    Building heights are computed from the predicted bounding boxes and building footprints, and subsequently, LoD1 building models are reconstructed. }
    } 
    \label{fig:net}
\end{figure*}

{We propose a footprint-guided bounding box regression network for building height retrieval that exploits the location relationship between a building's footprint and its bounding box.
Figure~\ref{fig:net} provides an overview of the proposed workflow. 
Specifically, in the network structure, we concatenate a SAR image and a building footprint mask as the input of the network. ResNet-101~\cite{he2016deep} is employed as the backbone. }
ResNet-101 has in total 101 weighted layers, including 5 blocks of convolutional layers, i.e., \textit{conv1}, \textit{conv2}, \textit{conv3}, \textit{conv4}, \textit{conv5}, and each contains a multi-layer deep subnetwork. 
First, \textit{conv1} to \textit{conv4} in ResNet are utilized to extract feature maps. 
{We extract the footprint bounding box from the building footprint mask and map it to the feature maps as the region of interest (RoI) of the building, i.e., the initial bounding box to be corrected. }
For each RoI, local features are pooled by RoI-Align~\cite{ren2015faster}. Then, \textit{conv5} of ResNet takes the pooled features, and a global average pooling layer and a fully connected layer proceed to predict corrections for the RoI with respect to the ground truth bounding box. The corrections are then added to the RoI of each building to produce its bounding box. Finally, building heights are derived from the predicted bounding boxes and are used to extrude LoD1 building models from the building footprint polygons.

For the parameterizations of bounding boxes, we adopt the $(x, y, w, h)$ coordinates used by R-CNN~\cite{girshick2014rich}. 
Let $\mathbf{B}=[x_B, y_B, w_B, h_B] \in \mathcal{R}^4 $ be the bounding box representation as a 4-dimensional vector, where $x$, $y$, $w$, and $h$ denote the box’s center coordinates and its width and height in an image patch. The task of bounding box regression is to regress a candidate bounding box $\mathbf{B}$ into a target bounding box $\mathbf{G}=[x_G, y_G, w_G, h_G]$. In our case, $\mathbf{B}$ is the footprint bounding box, and $\mathbf{G}$ is the building bounding box. 
%To encourage invariance to scale and location, t
The network predicts the distance vector $\bm{\Delta} = [\delta_x, \delta_y, \delta_w, \delta_h]$: 
\begin{equation}
    \begin{cases}
    \delta_x = (x_G - x_B)/w_B,\\
    \delta_y = (y_G - y_B)/h_B,\\
    \delta_w = log(w_G/w_B),\\
    \delta_h = log(h_G/h_B). 
    \end{cases}
\end{equation}

We employ the complete intersection over union (CIoU) loss~\cite{zheng2020distance}, which considers three geometric factors of bounding boxes: the overlap area, the central point distance, and the aspect ratio. CIoU is defined as: 
\begin{equation}
    \mathcal{L}_{CIoU}=1-IoU-\frac{\rho^2(\mathbf{b}, \mathbf{g})}{c^2}+\alpha v,
\end{equation}

where $\mathbf{b}$ and $\mathbf{g}$ denote the central points of $\mathbf{B}$ and $\mathbf{G}$, $\rho$ is the Euclidean distance, $c$ is the diagonal length of the smallest enclosing box covering the two boxes, $\alpha$ is a positive trade-off parameter, and $v$ measures the consistency of the aspect ratio. $IoU$, $\alpha$, and $v$ are defined as follows:
\begin{equation}
    \begin{split}
    IoU = \frac{|\mathbf{B} \cap \mathbf{G}|}{|\mathbf{B}\cup\mathbf{G}|}, \hspace{0.3cm}
    \alpha=\frac{v}{(1-IoU)+v}, \hspace{0.3cm}
    v=\frac{4}{\pi^2}(arctan\frac{w^{g}}{h^{g}}-arctan\frac{w^b}{h^b})^2. 
    \end{split}
\end{equation}
\section{Reference Data Generation}\label{sec:3}

For training our network, building bounding boxes as reference data and building footprints as input data in the SAR image coordinate system are necessary.   For this reason, we develop a workflow that employs building footprint and height data to automatically label building bounding boxes and building footprints in SAR images. The proposed workflow comprises three steps that are illustrated in Figure~\ref{fig:gt}: 
1) building heights acquisition, % from available data.  
2) building footprint masks generation, and 
3) building bounding boxes generation. 
In the following sections, we explain the details.

\subsection{Building height acquisition}

The first step is to collect building data. 
For each building, we collect the building height $h$ and the ground height $h_{ground}$, 
along with its footprint coordinates $(x,y)$. 

\begin{figure*}[!]
        \centering
        \subfloat[DEM~\cite{hirschmuller2008Stereo}]{\includegraphics[trim=0cm 0cm 0cm .5cm, clip=true,width= .33\textwidth]{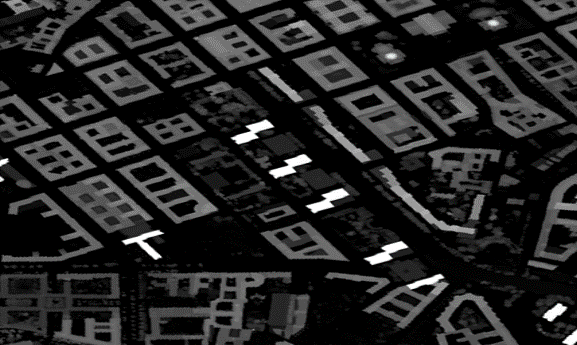}}\hfill
        \subfloat[LiDAR point clouds]{\includegraphics[trim=0cm .5cm 0cm 0cm, clip=true,width= 0.33\textwidth]{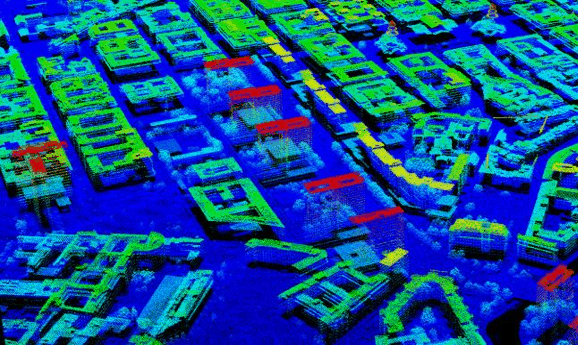}}\hfill
        \subfloat[Building models~\cite{Berlin3DDownloadportal}]{\includegraphics[trim=0cm .5cm 0cm 0cm, clip=true,width= 0.33\textwidth]{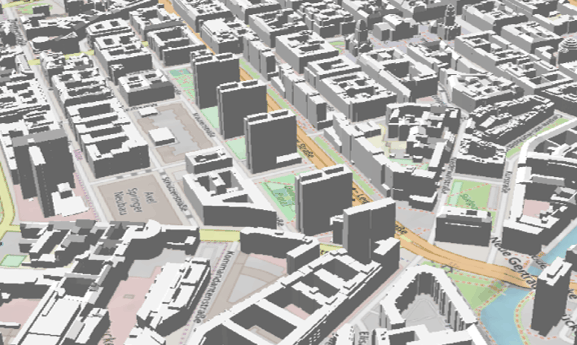}}
        \caption{Examples of {three different} building height sources in the same area in the city of Berlin, Germany.}
        \label{fig:heights}
    \end{figure*}

The proposed workflow only requires one single height value for one building, which can be acquired from various data types, such as city models, LiDAR data, and accurate DEMs. 
{Figure~\ref{fig:heights} shows an example of {three} data sources of building heights of the same area in Berlin.}
In some cities, public data sets are available that can be utilized to generate our annotations, 
e.g., Berlin city models~\cite{Berlin3DDownloadportal}, NYC open data~\cite{ny_ftp}, and 3D Buildings and Addresses of the Netherlands (3D BAG)~\cite{tudelft3d}. 
This loose requirement of the height data source significantly reduces the barrier of training data creation, which in turn supports the generation of reference data on a larger scale.

\subsection{Generation of building footprint masks in SAR images}   

In the previous step, building data are acquired in the UTM coordinate system. 
For our task, building footprints need to be projected to the SAR image coordinate system. That is to say, for each building footprint, its coordinates $(x,y,h_{ground})$ need to be transformed to $(rg, az)$, {where} $rg$ and $az$ denote range and azimuth coordinate, respectively. 
Generally, the coordinate transformation from the UTM coordinate system to the SAR imaging coordinate system includes iterative solving Doppler-Range-Ellipsoid equations that can be implemented with different approaches~\cite{4157311, schwabisch1998fast, toutin2004geometric, roth2004geocoding}. In this work, radar coding was performed using DLR's Integrated Wide Area Processor (IWAP)~\cite{gonzalez2013integrated}. Note that further registration is needed if the ground height $h_{ground}$ is not accurate~\cite{sun2020auto, sun2019automatic}. 

Then, building footprint masks are generated according to \textit{range-azimuth} coordinates of {the radar-coded vertices of building footprint polygons}. {For each building footprint mask, we set the pixel value to be 1 inside the footprint polygon and 0 elsewhere.}  

\begin{figure*}[h!]
    \centering
    \includegraphics[width= \linewidth]{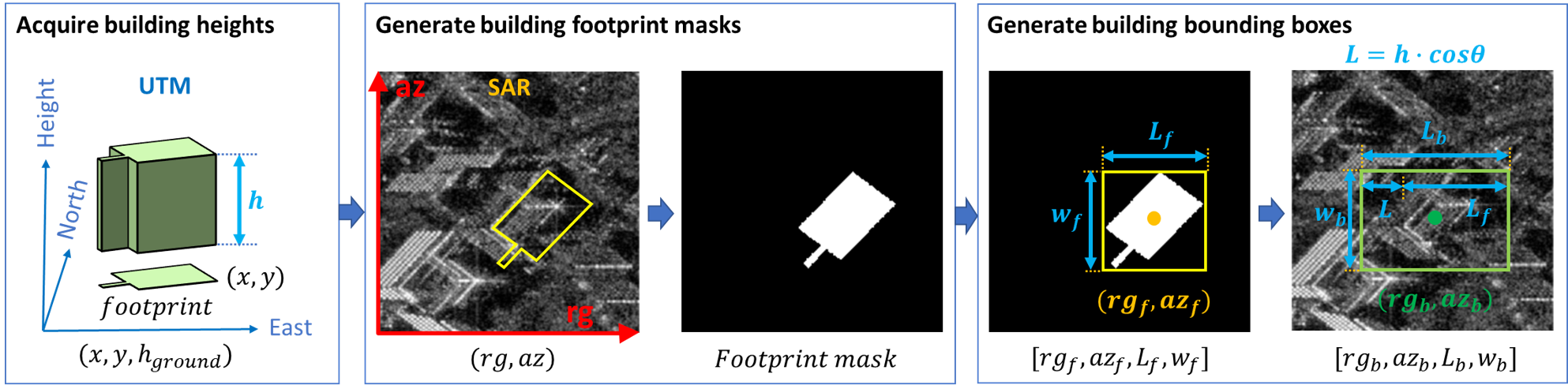}
    \caption{The workflow for dataset generation in three steps. 
    First, building footprints and height data are collected in the UTM coordinate system; then, they are projected to the SAR image coordinate system to generate building footprint masks; third, building bounding boxes are generated using footprint masks and building heights. 
    }
    \label{fig:gt}
\end{figure*}

\subsection{{Generation of the ground truth building bounding boxes in SAR images}}

{To generate the ground truth bounding box of a building, 
we first compute its footprint bounding box $B_f$. $B_f$ is defined by four values in pixels $[rg_f, az_f, L_f, w_f]$, in which $(rg_f, az_f)$ are coordinates of the %near-range/near-azimuth corner 
center point of the bounding box, and $L_f$ and $w_f$ are the width and height of the bounding box, respectively, as illustrated in Figure~\ref{fig:gt}. }

{Then, the building bounding box $B_b$ is generated from $B_f$. As illustrated in Figure~\ref{fig:b12_geo}, the difference between $B_b$ and $B_f$ results from the added width $L$, which is the layover length corresponding to the building height $h$:  
$L = h \cdot cos\theta$. %,  $L_b = L + L_f$. 
Therefore, the bounding box $B_b = [rg_b, az_b, L_b, w_b]$ can be generated} as: 
\begin{equation}
    \begin{cases}
    rg_b = rg_f - \frac{1}{2}L \\ 
    az_b = az_f\\
    L_b  = L + L_f\\
    w_b  = w_f 
    \end{cases}
\end{equation}

    Finally, we remove possible wrong bounding boxes. 
    Since the used SAR image and height data are often collected at different times, there might be inconsistencies resulted from urban changes, such as building construction and deconstruction. 
    We deal with this situation using the intensity values of the given SAR image. 
    In the SAR image, the intensity values are generally larger in building areas than in ground areas. 
    Therefore, a threshold is set to be the mode of the intensity values of the SAR image to exclude bounding boxes in which the mean intensity values are smaller than the threshold. 
\section{Experiments}\label{sec:test}

\subsection{Data description}

\begin{figure*}[h!]
    \centering
    \begin{tabular}{lr}
    \hspace{-0.2cm}
    \adjustbox{valign=b}{
    \subfloat[{Study area (blue region) in Berlin: the intersection of the spotlight SAR image (black rectangle)) and the DEM area (red rectangle).} ]{\includegraphics[width=.64\linewidth]{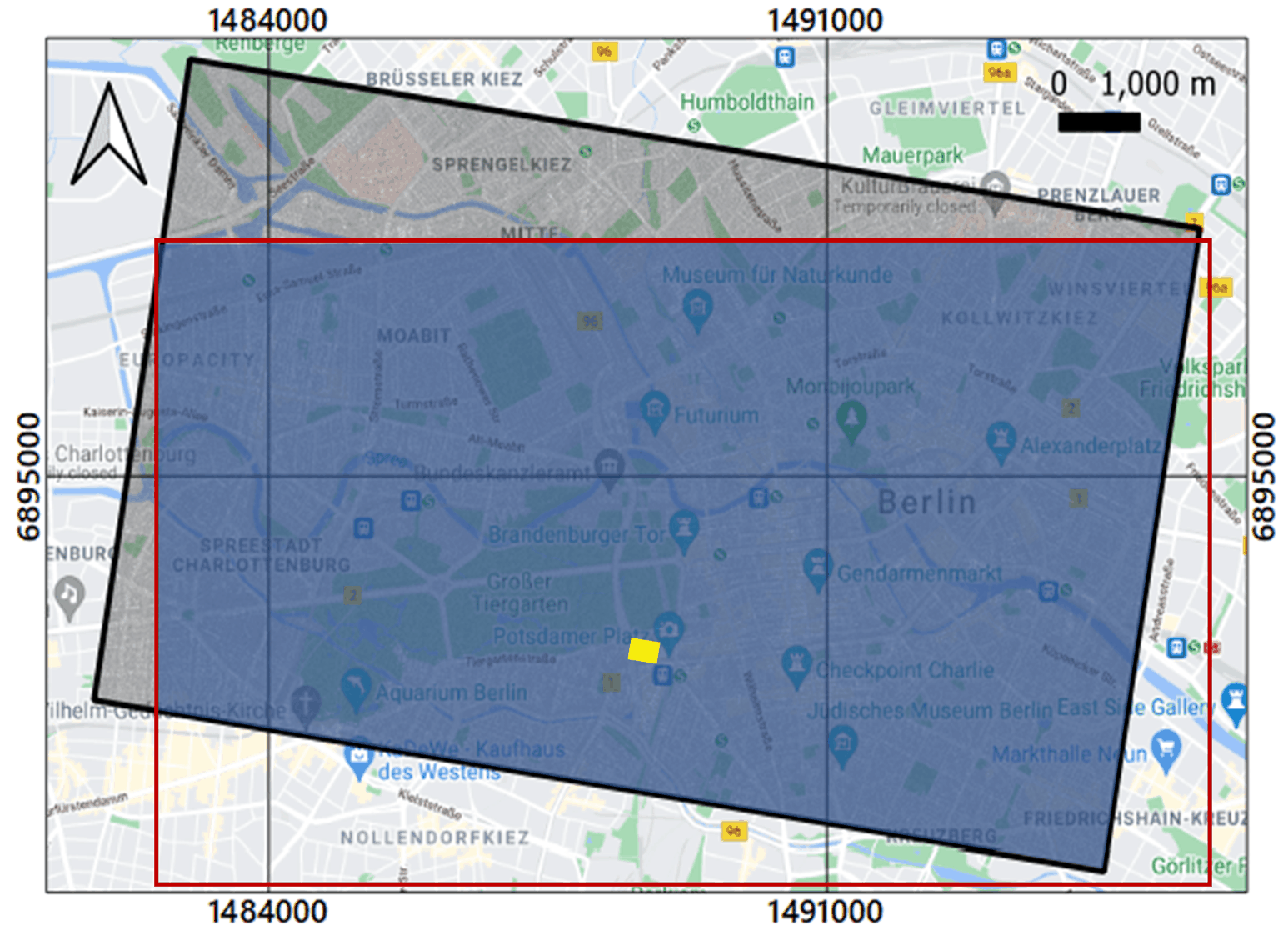}}}
    \hspace{0.1cm}
    \adjustbox{valign=b}{\begin{tabular}{@{}c@{}}
    \subfloat[Spotlight SAR image in the yellow rectangle in (a).]
    {\textcolor{yellow}{\fboxrule=3pt\fboxsep=0.2pt\fbox{\includegraphics[trim=0cm 1.cm 0cm 0cm,clip=true,width=.32\textwidth]{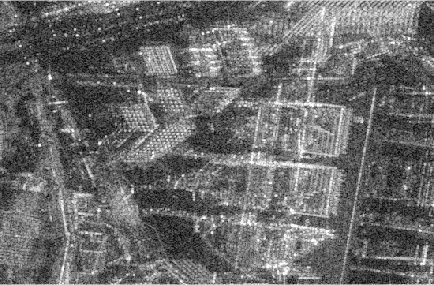}}}} \\
    \subfloat[Stripmap SAR image in the yellow rectangle in (a).]
    {\textcolor{yellow}{\fboxrule=3pt\fboxsep=0.2pt\fbox{{\includegraphics[trim=2cm 7.cm 1.9cm 8cm,clip=true,width=.32\textwidth]{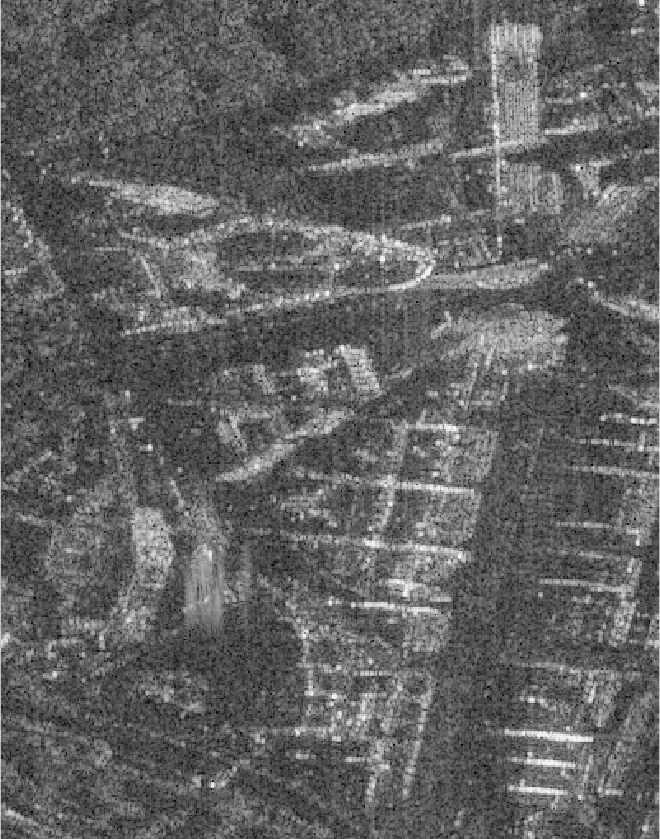}}}}}
    \end{tabular}}
    \end{tabular}
    \caption{The study area of both Berlin HS and Berlin SM data sets. (a) shows the area {(blue)} in the UTM coordinate system (UTM zone 32N). (b) and (c) show a comparison of the TerraSAR-X spotlight image and the stripmap image in the yellow rectangle in (a), respectively. }
\label{fig:areaberlin}
\end{figure*}

\begin{figure}[h!]
    \centering
    \subfloat{\includegraphics[width=.75\columnwidth]{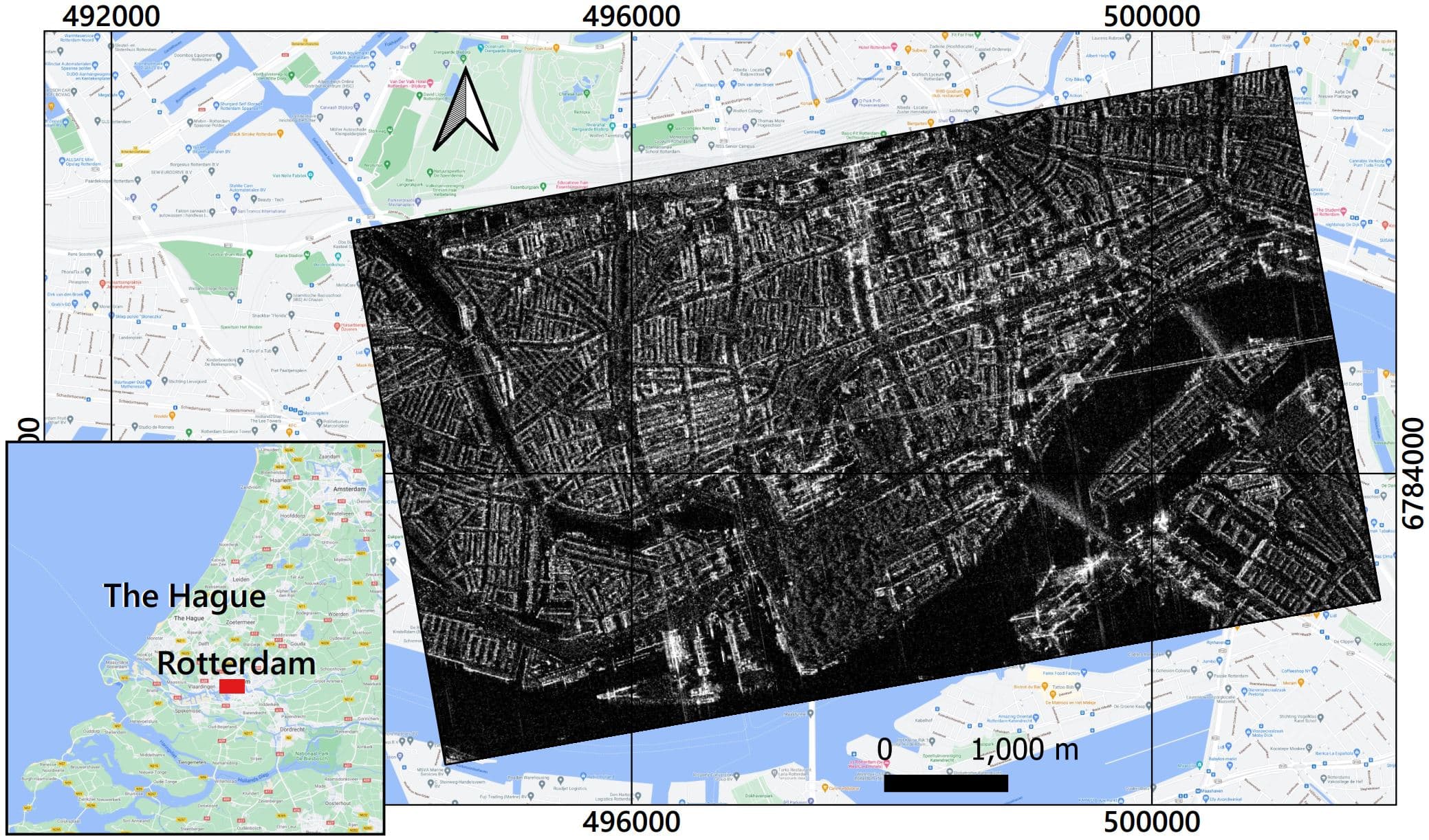}}
    \caption{The Rotterdam study area in the UTM coordinate system (UTM zone 31N).}
    \label{fig:area_rot}
\end{figure}

\begin{figure}[h!]
    \centering
    \subfloat{\includegraphics[width=.75\columnwidth]{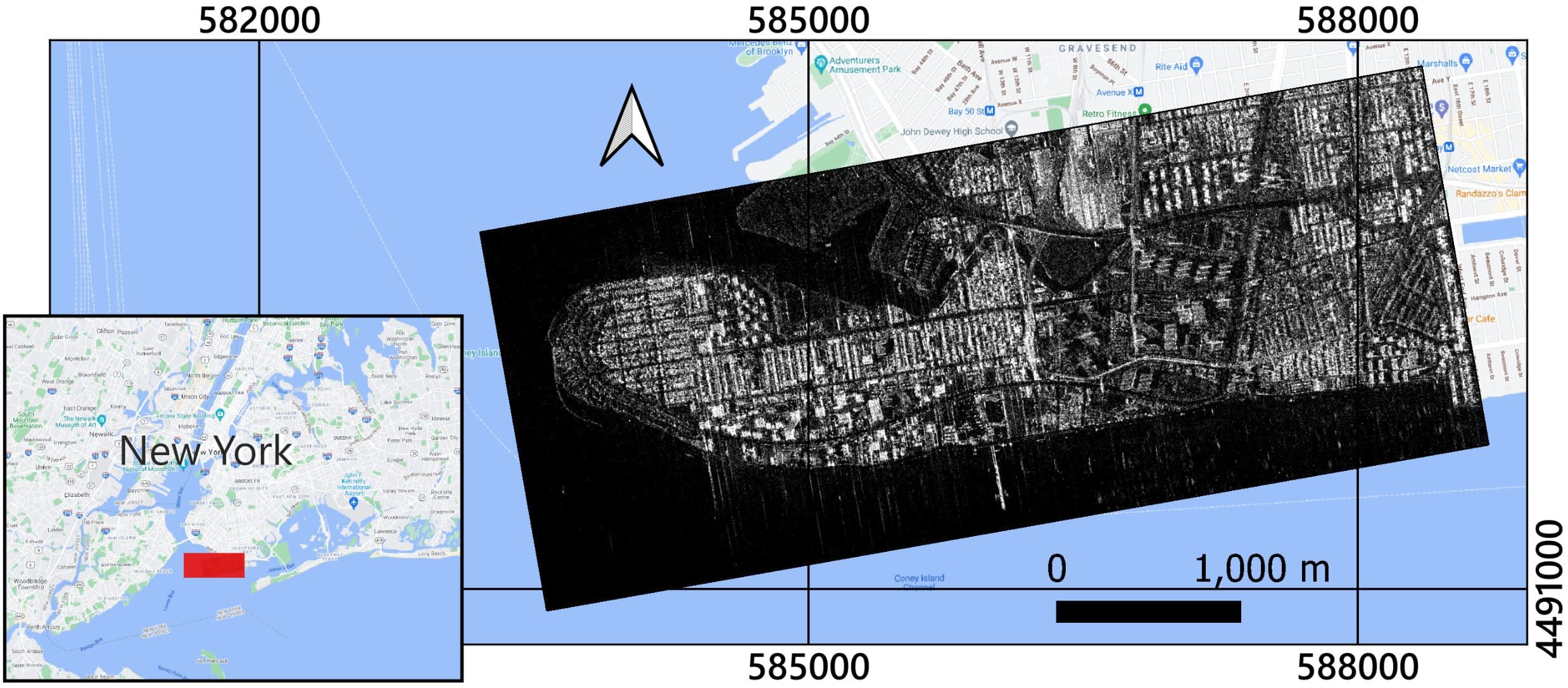}}
    \caption{The New York study area in the UTM coordinate system (UTM zone 18N). 
    }
    \label{fig:area_ny}
\end{figure}

The performance of the proposed method {is} evaluated on
four data sets, including one TerraSAR-X high-resolution spotlight (HS) image acquired over Berlin and three TerraSAR-X stripmap (SM) images acquired over Berlin, Rotterdam, and south Brooklyn in New York. Our four data sets are termed Berlin HS, Berlin SM, Rotterdam, and New York. 
Figure~\ref{fig:areaberlin} (a) shows the study region in Berlin, 
and the SAR images in Berlin HS and Berlin SM data sets are both cropped to cover the same region. 
Figure~\ref{fig:areaberlin} (b) and (c) show the spotlight image and the stripmap image in the yellow rectangle in (a), respectively.  
The study regions in Rotterdam and New York are shown in Figure~\ref{fig:area_rot} and Figure~\ref{fig:area_ny}, respectively. 

Table~\ref{tab:data} lists the main characteristics of the used SAR data, data sources of building footprints, and data sources of building heights in each data set. 
In this work, we make use of height data from LoD1 building models and accurate DEMs. LoD1 models represent buildings as blocks with flat roof structures and contain one height for each building~\cite{kolbe2005citygml}. As for DEMs, we regard the average roof height as the building height\footnote{\url{http://en.wiki.quality.sig3d.org/index.php/Modeling\_Guide\_for\_3D\_Objects \_-\_Part\_2:\_Modeling\_of\_Buildings\_(LoD1,\_LoD2,\_LoD3)}}. 
{By using the workflow described in Section~\ref{sec:3}, building bounding boxes and footprint masks are generated. }For each building, our data set contains a SAR image patch, a footprint mask, and a bounding box of the building.

\begin{table*}[h!]
\setlength{\tabcolsep}{2.5pt}
\small
    \centering
    \caption{{Characteristics of the used SAR data, data sources of building footprints, and data sources of heights in each data set.}}
    \label{tab:data}
    \begin{tabular}{lcccccc}
    \Xhline{2\arrayrulewidth}
    \multirow{ 2}{*}{Data set}  & TerraSAR-X  & pixel spacing: & pixel spacing: 
     & incidence & {data source:} & {data source:} \\
     & imaging mode & rg direction (m) & az direction (m) & angle ($^\circ$) & building footprints & building heights\\
     \hline
     \textbf{Berlin HS} & spotlight & 0.455  & 0.871  & 36.08 & Berlin 3D~\cite{Berlin3DDownloadportal} & DEM (7cm/pixel) \\
     \textbf{Berlin SM }& stripmap & 0.909  & 1.836  & 46.68 & Berlin 3D~\cite{Berlin3DDownloadportal} & DEM (7cm/pixel) \\
     \textbf{Rotterdam} & stripmap & 1.364  &1.852  & 39.28 & 3D BAG~\cite{tudelft3d} & 3D BAG~\cite{tudelft3d}\\
     \textbf{New York} & stripmap & 1.364  & 2.203  & 42.65 & NYC open data~\cite{ny_ftp} & NYC open data~\cite{ny_ftp}\\
     \Xhline{2\arrayrulewidth}
    \end{tabular}
\end{table*}

\subsection{Training details} \label{sec:train_details}

To train an effective and robust network, we first cropped the SAR image into patches. 
Patches containing incomplete footprints or bounding boxes are discarded. In the four data sets, the high-resolution spotlight SAR images are cropped into patches of $256 \times 256$ pixels with a stride of $150$ pixels,  and the stripmap SAR images are cropped into patches of $128 \times 128$ pixels with a stride of $70$ pixels. % considering the image resolution, 
Consequently, building data in the study areas are prepared, and each building has a ground truth bounding box and two patches: a SAR image patch and a footprint mask patch. All building samples are then divided to build the training set and the testing set. {We made sure that} the training and test regions do not overlap. 
{The network takes one SAR patch and the corresponding GIS patch for one building as inputs.}
Table~\ref{tab:trainTest} lists the patch size and sample numbers of training/testing sets of each data set. Before feeding data into models, 
data sets were normalized into the range of [0, 1]. 
For data sets generated using stripmap SAR images, image patches are re-scaled to $256 \times 256$ pixels. 

The network is implemented on PyTorch and trained on one NVIDIA Tesla P100 16GB GPU. During the training procedure,  
the layers of ResNet are initialized with weights pre-trained on ImageNet~\cite{imagenet_cvpr09}, 
and other layers are randomly initialized by drawing weights from a zero-mean Gaussian distribution with a standard deviation of 0.01. %The loss is defined as CIoU loss~\cite{zheng2020distance}. \todoyw{mentioned before} 
All weights are updated through back-propagation, and we use stochastic gradient descent (SGD)~\cite{zhang2014deep} as the optimizer.  
The learning rate is initialized as 0.001 and reduced by a factor of $0.1$ once the loss stops to decrease for three epochs. We use a momentum of 0.9 and a weight decay of 0.0005. 
{For hyperparameter settings of batch size and training epochs, we tested using the Berlin HS data set and the numerical results are shown in Table~\ref{tab:batch} and Table~\ref{tab:epochs}. The network is trained for 10 epochs, and we utilize a small batch size of 4. }

\begin{table}[h!]
\small
\centering
    \caption{{Numerical results on Berlin HS data set for choosing the batch size. All experiments are trained for 10 epochs. The highest values of the metrics are highlighted in {\textbf{bold}}.}}
    \label{tab:batch}
    \begin{tabular}{ccccc}
    \Xhline{2\arrayrulewidth}
    %  & patch size  &  cropping stride & total number of samples & training samples & test samples \\
    {batch size} & 4  &  16 & 64  & 256  \\
     \hline
     $he_{mean}$ (m) & \textbf{4.3} & 4.6  & 4.9 & 5.3 \\
     $he_{std}$ (m) &  \textbf{6.3} & 6.6  & 5.8 & 7.2 \\
     \Xhline{2\arrayrulewidth}
    \end{tabular}
\end{table}

\begin{table}[h!]
\small
\centering
    \caption{{Numerical results on Berlin HS data set for choosing the training epochs. All experiments use the batch size of 4. The highest values of the metrics are highlighted in {\textbf{bold}}.}}
    \label{tab:epochs}
    \begin{tabular}{ccccccccccc}
    \Xhline{2\arrayrulewidth}
    %  & patch size  &  cropping stride & total number of samples & training samples & test samples \\
    {training epochs} &  5   & 6    &  7   & 8    & 9    & \textbf{10} & 11& 12 & 15 & 20 \\
     \hline
     $he_{mean}$ (m) & 4.7 & 4.4  & 4.5 & 4.4 & 4.4 & \textbf{4.3}& {4.3} &4.3 &  4.3 & 4.3\\
     $he_{std}$ (m)  & 6.8 & 6.6  & 6.5 & 6.4 & 6.4 & \textbf{6.3}& {6.4} &6.3 &  6.3 & 6.4\\
     \Xhline{2\arrayrulewidth}
    \end{tabular}
\end{table}

\begin{table}[h!]
\small
\centering
    \caption{{Patch size and sample numbers in each data set. }}
    \label{tab:trainTest}
    \begin{tabular}{lccccc}
    \Xhline{2\arrayrulewidth}
    %  & patch size  &  cropping stride & total number of samples & training samples & test samples \\
    \multirow{ 2}{*}{Data set} & patch size  &  cropping & total  & training  & testing \\
     & (pixel)  &  stride &  samples &  samples &  samples \\
     \hline
     \textbf{Berlin HS} & $256\times256$ & 150  & 29842  & 19251 & 10591 \\
     \textbf{Berlin SM }& $128\times128$ & 70 & 17184 & 15863 & 1321\\
     \textbf{Rotterdam} & $128\times128$ & 70 & 15054  & 13368 & 1686 \\
     \textbf{New York} & $128\times128$ & 70 & 7922 & 7318 & 604\\
     \Xhline{2\arrayrulewidth}
    \end{tabular}
\end{table}

\subsection{Comparative experiments}\label{sec:comp_ex}

For our problem, the major focus is to predict the bounding box of each building correctly. 
As bounding box regression is also an important task for object detection, object detection networks can be employed for our problem by deriving building heights from the predicted bounding boxes.  

In the experiments, we utilize five object detection models to estimate building heights and compare their results with ours.
The object detection networks include three one-stage networks, SSD~\cite{liu2016ssd}, YOLOv3~\cite{redmon2018yolov3}, RetinaNet~\cite{lin2017focal}, 
and a two-stage network, Faster R-CNN~\cite{ren2015faster}. 
Additionally, feature pyramid network (FPN)~\cite{lin2017feature} is combined with Faster R-CNN in its backbone, termed as {Faster R-CNN w. FPN}, for better detecting objects at different scales. 
We denote the combined procedures of object detection and height estimation as SSD$_h$, YOLOv3$_h$, RetinaNet$_h$, Faster R-CNN$_h$, Faster R-CNN w. FPN$_h$.

For implementation, MMdetection~\cite{mmdetection} is employed for SSD$_h$, YOLOv3$_h$, RetinaNet$_h$, and Faster R-CNN w. FPN$_h$, 
and the implementation in~\cite{jjfaster2rcnn} is utilized for Faster R-CNN$_h$. 
ResNet-101 is used as the backbone for RetinaNet$_h$, Faster R-CNN$_h$, and Faster R-CNN w. FPN$_h$. 
For all the networks, the input is the concatenated SAR image and the building footprint mask. All input image patches are re-scaled to $256 \times 256$ pixels. 
Other default parameters in each implementation are kept.

\subsection{Quantitative evaluation}

The performance of networks is evaluated based on two criteria: the height accuracy and training time. 
We record the training time that each model takes for training on each data set and calculate building heights from the predicted bounding boxes, as stated in Section~\ref{sec:method}. 
The accuracy of retrieved building heights is measured by the mean ($he_{mean}$) and standard deviation ($he_{std}$) of height errors of all buildings $H_e$: 
\begin{equation}
\begin{split}
    & he_{mean} = mean(H_e), \\
    & he_{std} = std(H_e).  
\end{split}
\end{equation}
$H_e = \{{h^i_{true} - h^i_{predict} | i=1,...,n}\}$, where ${h^i_{true}}$ and $h^i_{predict}$ are the ground truth height and predicted height for the $i$-th building, respectively, and $n$ is the number of test samples.

\begin{figure*}[h!]
    \centering
    \subfloat[{Berlin HS}]{\includegraphics[width= .33\textwidth]{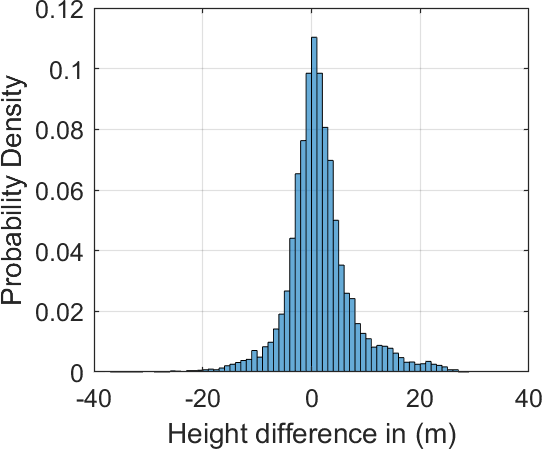}}
    \hspace{2cm}
    \subfloat[Berlin SM]{\includegraphics[width= .33\textwidth]{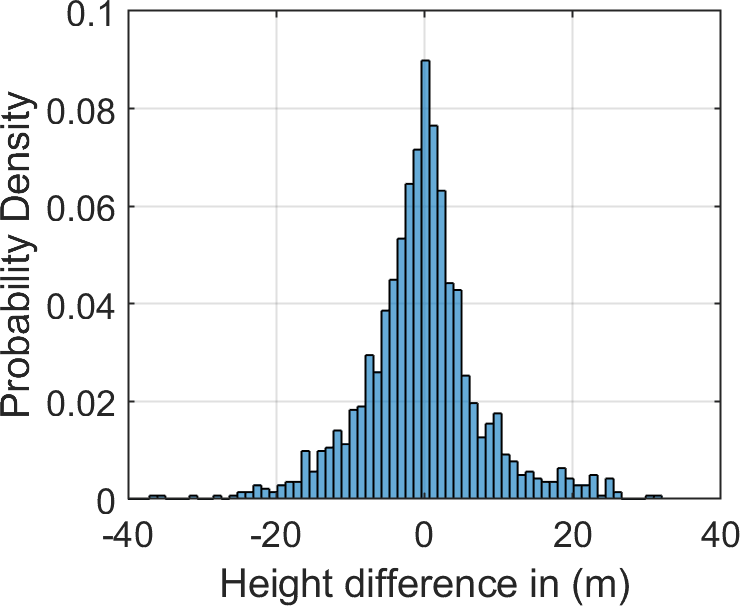}}
    \\
    \subfloat[Rotterdam]{\includegraphics[width= .33\textwidth]{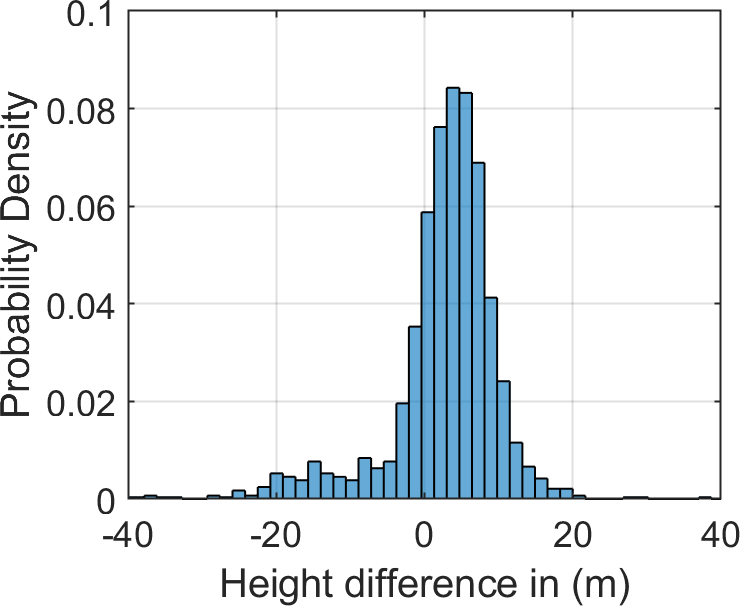}}
    \hspace{2cm}
    \subfloat[New York]{\includegraphics[width= .33\textwidth]{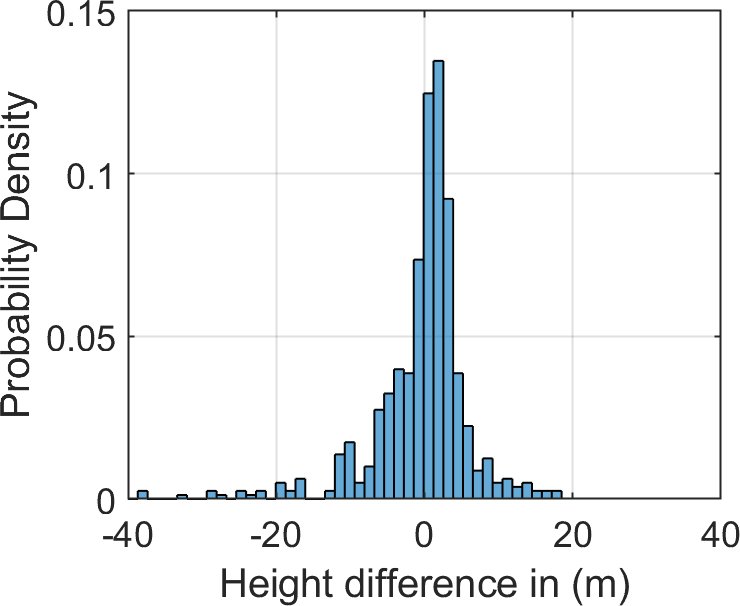}}
    \caption{Histogram of building height errors predicted with our network in the study areas.}
    \label{fig:he_Hist}
\end{figure*}

Table~\ref{tab:num} reports numerical results of different models on four data sets, and Figure~\ref{fig:he_Hist} shows histograms of height errors predicted by our network. 
We observe that Faster R-CNN$_h$ performs the best in all four data sets among all the networks. However, the results are only 0.1-0.2 m better than those achieved by our network, which are trivial comparing to the absolute height errors (in the range of 4.3 m to 5.6 m). 
The results show that our networks, RetinaNet$_h$, and Faster R-CNN$_h$, outperform SSD$_h$ and YOLOv3$_h$ in height accuracy. 
Interestingly, FPN did not bring improvement to Faster R-CNN$_h$. One reason could be that the difference in the scale of building footprints is not particularly large. 

On Berlin HS data set, all networks achieve the best performance in terms of height accuracy comparing to other data sets, owing to the higher spatial resolution of the spotlight image than the stripmap images. 
However, we notice that the differences are not significant.  
For instance, 
using stripmap images, the mean height error achieved by Faster R-CNN$_h$ ranges from 4.7 m to 5.6 m, and the standard deviation from 7.1 m to 7.6 m, depending on the data set. 
While using spotlight data (Berlin HS), the mean height error achieved is 4.3 m, and the standard deviation is 6.4 m, which is only 1-1.3 m better than those achieved using stripmap images.

In terms of the speed, 
our method significantly outperforms not only two-stage networks such as Faster R-CNN$_h$ but also fast networks like SSD$_h$ and YOLOv3$_h$. 
Comparing to Faster R-CNN$_h$, the training time of our network reduces about 80\%. 
The computation of the networks is reduced owing to the utilization of footprint bounding boxes. The fast training speed is particularly important when working with large data sets. 

{Our network outperforms the detection-based networks mainly due to the tailored use of building footprints, i.e., the module is designed to extract the footprint bounding box as the initial bounding box specifically for our task. The detection networks, on the other hand, lack the module specified for extracting building footprint information. They rely on a large amount of region proposals to obtain possible initial bounding boxes. In addition, our network provides one initial proposal, i.e., footprint bounding box, for each bounding box. However, the detection networks must provide multiple proposals in the earlier stage and rely on the classification scores to select the final bounding box in the later stage. Therefore, the computational cost of our network is much smaller. }

To sum up, our network achieves accuracy comparable with Faster R-CNN$_h$ and much superior performance on speed by effectively using the multi-modal information contained in GIS data. 
The comparison of these results corroborates that the proposed network can significantly reduce the computational cost while keeping the height accuracy.

\begin{table*}[!]
\small
    \centering
    \caption{{
    Numerical results on four data sets. The highest values of different metrics are highlighted in {\textbf{bold}}. }
    }\label{tab:num}
    \begin{tabular}{llccc}
    \Xhline{2\arrayrulewidth}
    Data set &  Model Name       & $he_{mean}$ (m)  & $he_{std}$ (m) &   Training Time \\
    \hline
    \multirow{6}{*}{\textbf{Berlin HS}}   
        &{SSD$_h$ }                        &{6.6} & {9.4}   &          {3h26mins} \\ 
        &{YOLOv3$_h$  }                    & {6.0} & {8.1}   &          {4h16mins} \\
        &{RetinaNet$_h$}               & {4.7 }& {6.5}  &           {5h22mins} \\
        &{Faster R-CNN w.FPN$_h$    }      & {5.0 }& {7.3  } &           {5h10mins}\\
        &{Faster R-CNN$_h$   }             & {\textbf{4.3} }& {\textbf{6.2}} & {5h26mins} \\
        &{Ours }                        & {\textbf{4.3} }& {6.3}   &        {\textbf{1h01mins}}\\
    \hline
    \multirow{6}{*}{\textbf{Berlin SM}}   
        & SSD$_h$                        & 7.9  & 10.3   &         1h59mins\\ 
        & YOLOv3$_h$                    & 6.5  & 9.8    &         2h22mins \\
        &RetinaNet$_h$                   & 5.9  & 9.0    &         3h32mins \\
        &Faster R-CNN w.FPN$_h$          & 6.1  & 8.7    &         3h25mins\\
        &Faster R-CNN$_h$                & \textbf{5.6}  & \textbf{7.1} & 3h28mins\\
        &Ours                         & 5.7  & 7.2    &  \textbf{52mins} \\
    \hline
    \multirow{6}{*}{\textbf{Rotterdam}}   
        & SSD$_h$                     & 6.4 & 9.5  &           1h47mins\\ 
        &YOLOv3$_h$                  & 5.9 & 8.3  &           2h13mins \\
        &RetinaNet$_h$                & 5.4 & 7.6  &           3h23mins \\
        &Faster R-CNN w.FPN$_h$       & 5.8 & 7.8  &           3h14mins\\
        &Faster R-CNN$_h$             & \textbf{5.4} & \textbf{7.6} & 3h40mins\\
        &Ours                      & 5.5 & \textbf{7.6}  &          {\textbf{44mins}}\\
    \hline
    \multirow{6}{*}{\textbf{New York}}   
        & SSD$_h$                     & 6.2 & 12.2   &             57mins\\ 
        &YOLOv3$_h$                  & 6.2 & 13.2   &             1h15mins \\
        &RetinaNet$_h$                & 4.8 & 7.3    &             1h55mins\\
        &Faster R-CNN w.FPN$_h$       & 5.0 & 7.8    &             1h30mins\\
        &Faster R-CNN$_h$             & \textbf{4.7} & \textbf{7.3} & 1h59mins \\
        &Ours                      & 4.9 & 7.6    &             \textbf{26mins}\\
    \Xhline{2\arrayrulewidth}
    \end{tabular}
\end{table*}

\subsection{Qualitative evaluation}

\newlength{\tempdima}
\newcommand{\rowname}[1]% #1 = text
{\rotatebox{90}{\makebox[\tempdima][c]{{#1}}}}

\renewcommand{\thesubfigure}{\alph{subfigure}}
\newcommand{\mycaption}[1]% #1 = caption
{\refstepcounter{subfigure}{(\thesubfigure) }{\ignorespaces #1}}

\begin{figure*}[!]
    \centering
    \settoheight{\tempdima}{\includegraphics[width=.08\textwidth]{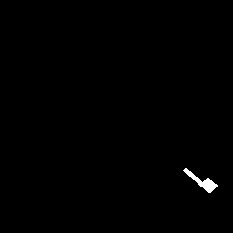}}%
    \begin{tabularx}{0.99\linewidth}{@{\hskip3pt}c@{\hskip3pt}c@{\hskip3pt}c@{\hskip3pt}c@{\hskip3pt}c@{\hskip3pt}c@{\hskip3pt}c@{\hskip3pt}}
    \\\vspace{-0.35cm}\hspace{0.3cm}
      \hspace{.3cm}{\footnotesize {b1 (Berlin HS)}} & {\footnotesize {b2 (Berlin HS)}} 
    & {\footnotesize {b3 (Berlin HS)}} & {\footnotesize {b4 (Berlin SM)}} 
    & {\footnotesize {b5 (Berlin SM)}} & {\footnotesize {b6 (Berlin SM)}}
    \\\vspace{-0.4cm}\hspace{0.3cm}
    \rowname{\hspace{1.1cm}\scriptsize{Footprint mask}}
    \subfloat{{\includegraphics[width=.142\textwidth]{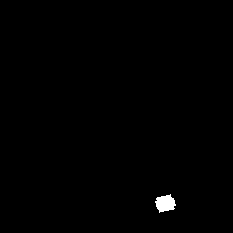}}}&
    \subfloat{{\includegraphics[width=.142\textwidth]{figs/berlinHS/00201_1_gis1.png}}}& 
    \subfloat{{\includegraphics[width=.142\textwidth]{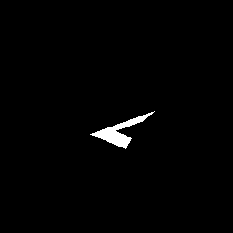}}}&
    \subfloat{{\includegraphics[width=.142\textwidth]{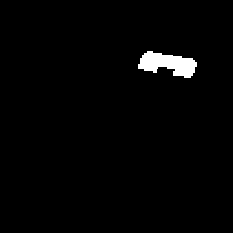}}}&
    \subfloat{{\includegraphics[width=.142\textwidth]{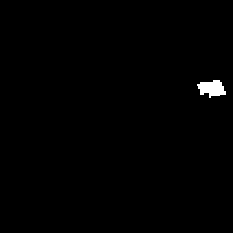}}}&
    \subfloat{{\includegraphics[width=.142\textwidth]{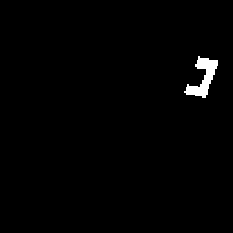}}}&
    \\\vspace{-0.4cm}\hspace{0.3cm}
    \rowname{\hspace{1.1cm}\scriptsize{SAR image}}
    \subfloat{{\includegraphics[width=.142\textwidth]{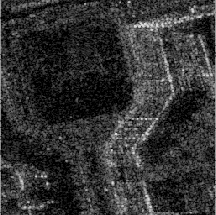}}}&
    \subfloat{{\includegraphics[width=.142\textwidth]{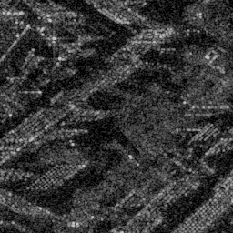}}}& 
    \subfloat{{\includegraphics[width=.142\textwidth]{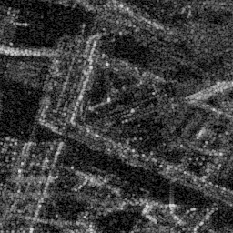}}}&
    \subfloat{{\includegraphics[width=.142\textwidth]{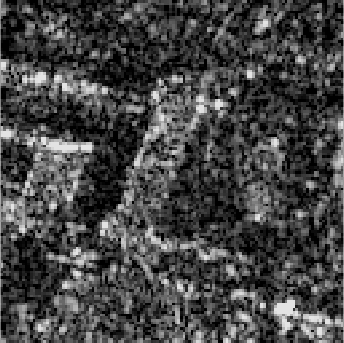}}}&
    \subfloat{{\includegraphics[width=.142\textwidth]{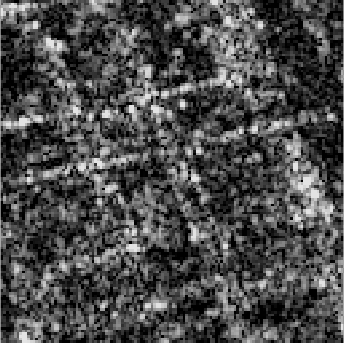}}}&
    \subfloat{{\includegraphics[width=.142\textwidth]{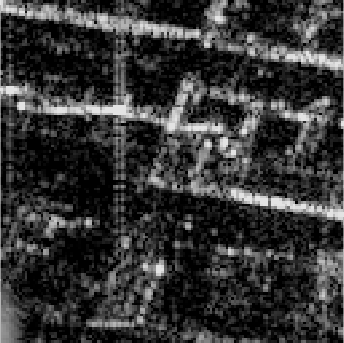}}}&
    \\\vspace{-0.4cm}\hspace{0.3cm}
    \rowname{\hspace{1.1cm}\scriptsize{SSD$_h$}}
    \subfloat{{\includegraphics[width=.142\textwidth]{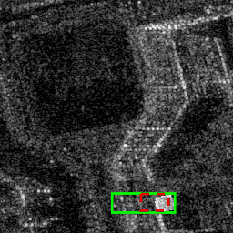}}}&
    \subfloat{{\includegraphics[width=.142\textwidth]{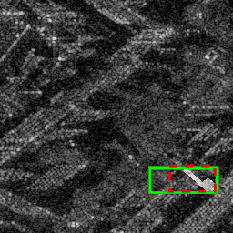}}}& 
    \subfloat{{\includegraphics[width=.142\textwidth]{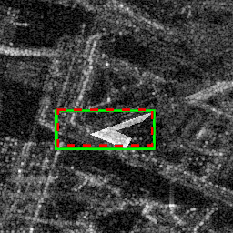}}}&
    \subfloat{{\includegraphics[width=.142\textwidth]{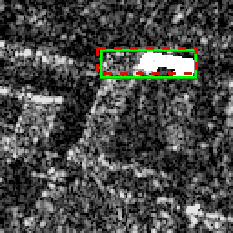}}}&
    \subfloat{{\includegraphics[width=.142\textwidth]{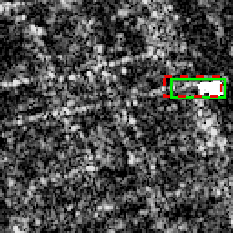}}}&
    \subfloat{{\includegraphics[width=.142\textwidth]{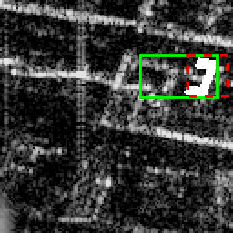}}}&
    \\\vspace{-0.4cm}\hspace{0.3cm}
    \rowname{\hspace{1.1cm}\scriptsize{YOLOv3$_h$}} 
    \subfloat{{\includegraphics[width=.142\textwidth]{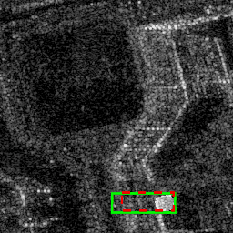}}}&
    \subfloat{{\includegraphics[width=.142\textwidth]{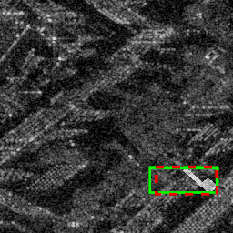}}}& 
    \subfloat{{\includegraphics[width=.142\textwidth]{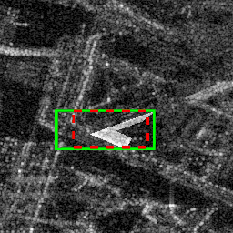}}}&
    \subfloat{{\includegraphics[width=.142\textwidth]{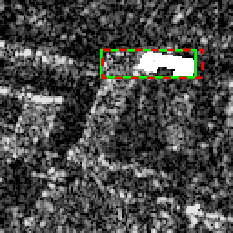}}}&
    \subfloat{{\includegraphics[width=.142\textwidth]{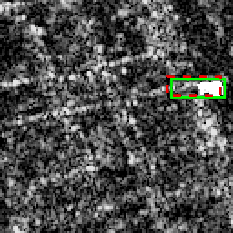}}}&
    \subfloat{{\includegraphics[width=.142\textwidth]{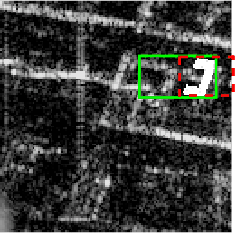}}}&
    \\\vspace{-0.4cm}\hspace{0.3cm}
    \rowname{\hspace{1.1cm}\scriptsize{RetinaNet$_h$}} 
    \subfloat{{\includegraphics[width=.142\textwidth]{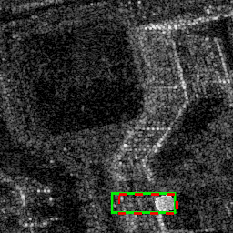}}}&
    \subfloat{{\includegraphics[width=.142\textwidth]{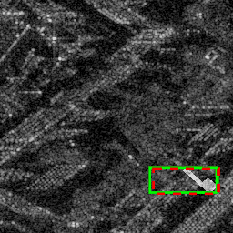}}}& 
    \subfloat{{\includegraphics[width=.142\textwidth]{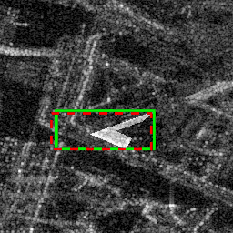}}}&
    \subfloat{{\includegraphics[width=.142\textwidth]{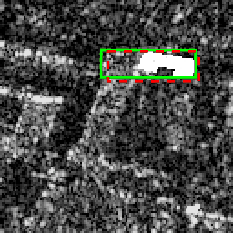}}}&
    \subfloat{{\includegraphics[width=.142\textwidth]{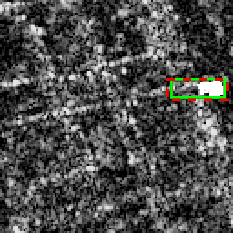}}}&
    \subfloat{{\includegraphics[width=.142\textwidth]{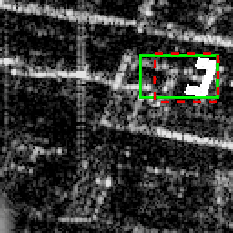}}}&
    \\\vspace{-0.4cm}\hspace{0.3cm}
    \rowname{\hspace{1.1cm}\scriptsize{Faster R-CNN$_h$}} 
    \subfloat{{\includegraphics[width=.142\textwidth]{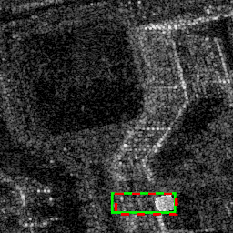}}}&
    \subfloat{{\includegraphics[width=.142\textwidth]{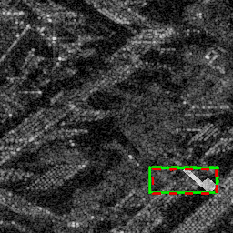}}}& 
    \subfloat{{\includegraphics[width=.142\textwidth]{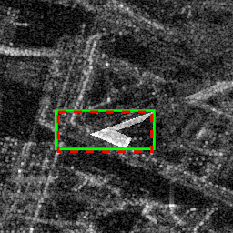}}}&
    \subfloat{{\includegraphics[width=.142\textwidth]{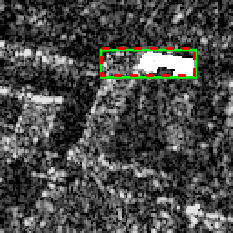}}}&
    \subfloat{{\includegraphics[width=.142\textwidth]{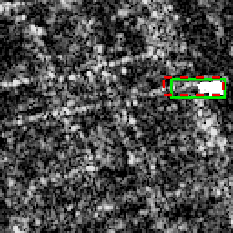}}}&
    \subfloat{{\includegraphics[width=.142\textwidth]{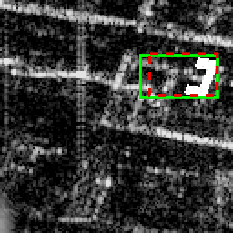}}}&
    \\\vspace{-0.4cm}\hspace{0.3cm}
    \rowname{\hspace{1.cm}\tiny{Faster R-CNN w.FPN$_h$}} 
    \subfloat{{\includegraphics[width=.142\textwidth]{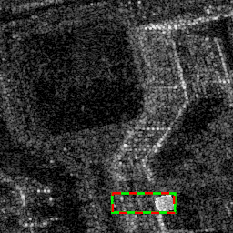}}}&
    \subfloat{{\includegraphics[width=.142\textwidth]{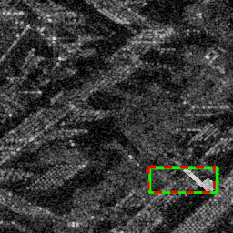}}}& 
    \subfloat{{\includegraphics[width=.142\textwidth]{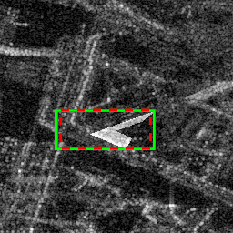}}}&
    \subfloat{{\includegraphics[width=.142\textwidth]{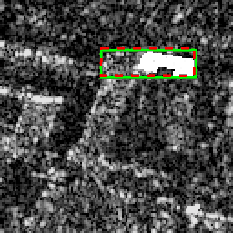}}}&
    \subfloat{{\includegraphics[width=.142\textwidth]{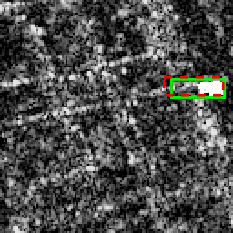}}}&
    \subfloat{{\includegraphics[width=.142\textwidth]{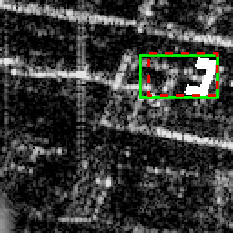}}}&
    \\\vspace{-0.4cm}\hspace{0.3cm}
    \rowname{\hspace{1.1cm}\scriptsize{Ours}} 
    \subfloat{{\includegraphics[width=.142\textwidth]{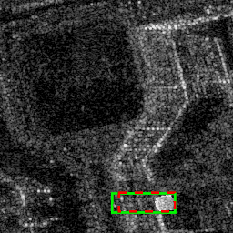}}}&
    \subfloat{{\includegraphics[width=.142\textwidth]{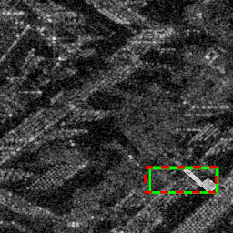}}}& 
    \subfloat{{\includegraphics[width=.142\textwidth]{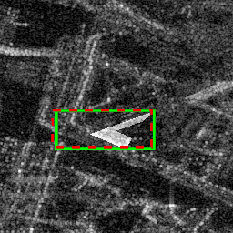}}}&
    \subfloat{{\includegraphics[width=.142\textwidth]{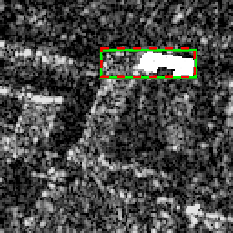}}}&
    \subfloat{{\includegraphics[width=.142\textwidth]{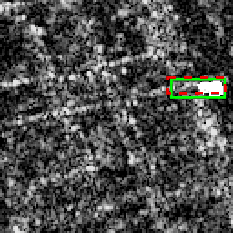}}}&
    \subfloat{{\includegraphics[width=.142\textwidth]{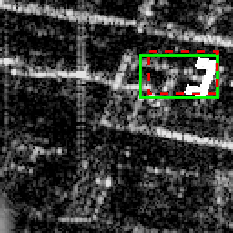}}}
    \vspace{0.22cm}
    \end{tabularx}
    \caption{Examples of predicted bounding boxes using different networks in Berlin HS and Berlin SM datasets. The predicted and ground truth bounding boxes are marked in red and green, respectively. 
    {The ground truth bounding boxes are obtained using the procedure described in Section 3.3. }
    %\YW{$>>$caption not correspond to the figure. Text are overlapping on the left of the figure. Please fix. Also check the next figure.$<<$} 
    }   \label{fig:subgisF}
\end{figure*}

\begin{figure*}[!]
    \centering
    \settoheight{\tempdima}{\includegraphics[width=.08\textwidth]{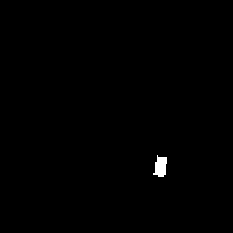}}%
    \begin{tabularx}{0.99\linewidth}{@{\hskip3pt}c@{\hskip3pt}c@{\hskip3pt}c@{\hskip3pt}c@{\hskip3pt}c@{\hskip3pt}c@{\hskip3pt}c@{\hskip3pt}}
    \\\vspace{-0.35cm}\hspace{0.3cm}
    \hspace{0.3cm}{\footnotesize {b7 (Rotterdam)}} & {\footnotesize {b8 (Rotterdam)}} & {\footnotesize {b9 (Rotterdam)}} & {\footnotesize {b10 (New York)}} & {\footnotesize {b11 (New York)}} & {\footnotesize {b12 (New York)}}
    \\\vspace{-0.4cm}\hspace{0.3cm}
    \rowname{\hspace{1.1cm}\scriptsize{Footprint mask}}
    \subfloat{{\includegraphics[width=.142\textwidth]{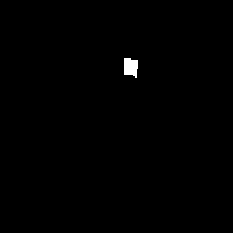}}}&
    \subfloat{{\includegraphics[width=.142\textwidth]{figs/Rot/00461_4_gis1.png}}}& 
    \subfloat{{\includegraphics[width=.142\textwidth]{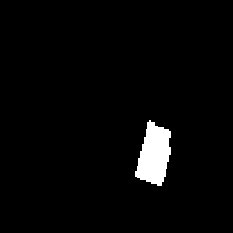}}}&
    \subfloat{{\includegraphics[width=.142\textwidth]{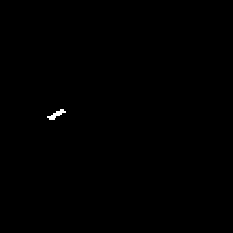}}}&
    \subfloat{{\includegraphics[width=.142\textwidth]{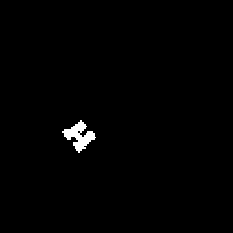}}}&
    \subfloat{{\includegraphics[width=.142\textwidth]{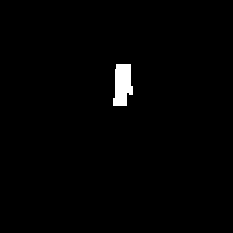}}}&
    \\\vspace{-0.4cm}\hspace{0.3cm}
    \rowname{\hspace{1.1cm}\scriptsize{SAR image}}
    \subfloat{{\includegraphics[width=.142\textwidth]{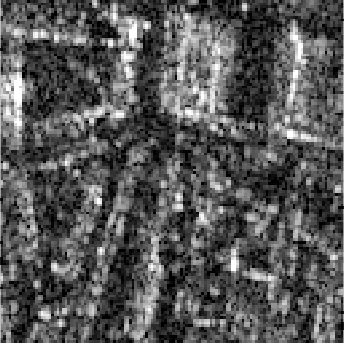}}}&
    \subfloat{{\includegraphics[width=.142\textwidth]{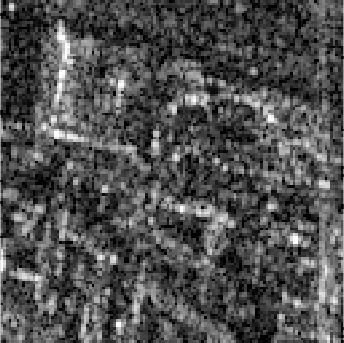}}}& 
    \subfloat{{\includegraphics[width=.142\textwidth]{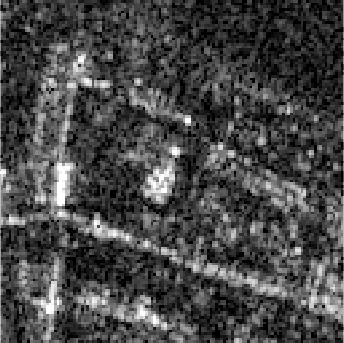}}}&
    \subfloat{{\includegraphics[width=.142\textwidth]{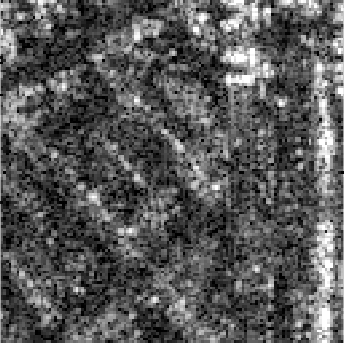}}}&
    \subfloat{{\includegraphics[width=.142\textwidth]{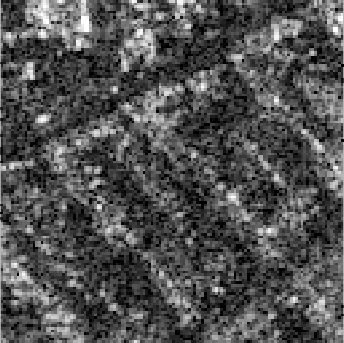}}}&
    \subfloat{{\includegraphics[width=.142\textwidth]{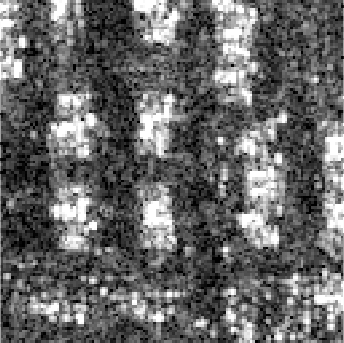}}}&
    \\\vspace{-0.4cm}\hspace{0.3cm}
    \rowname{\hspace{1.1cm}\scriptsize{SSD$_h$}}
    \subfloat{{\includegraphics[width=.142\textwidth]{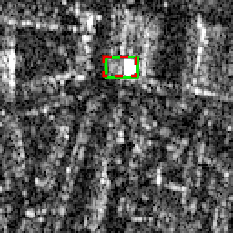}}}&
    \subfloat{{\includegraphics[width=.142\textwidth]{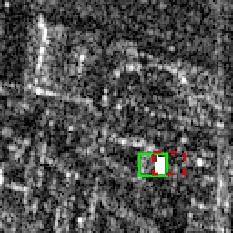}}}& 
    \subfloat{{\includegraphics[width=.142\textwidth]{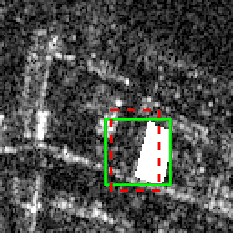}}}&
    \subfloat{{\includegraphics[width=.142\textwidth]{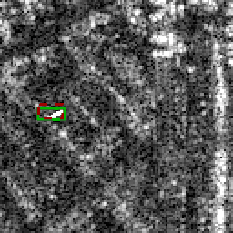}}}&
    \subfloat{{\includegraphics[width=.142\textwidth]{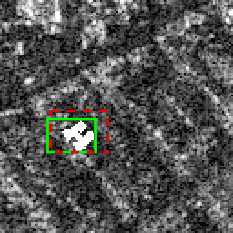}}}&
    \subfloat{{\includegraphics[width=.142\textwidth]{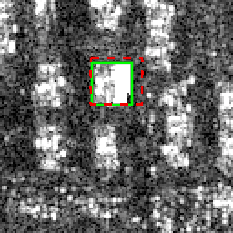}}}&
    \\\vspace{-0.4cm}\hspace{0.3cm}
    \rowname{\hspace{1.1cm}\scriptsize{YOLOv3$_h$}} 
    \subfloat{{\includegraphics[width=.142\textwidth]{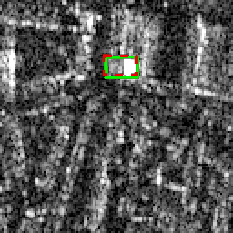}}}&
    \subfloat{{\includegraphics[width=.142\textwidth]{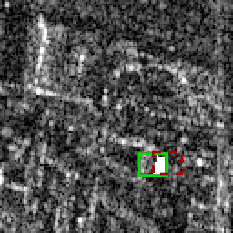}}}& 
    \subfloat{{\includegraphics[width=.142\textwidth]{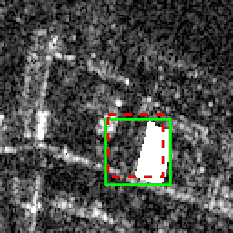}}}&
    \subfloat{{\includegraphics[width=.142\textwidth]{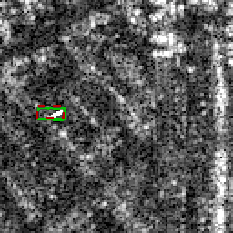}}}&
    \subfloat{{\includegraphics[width=.142\textwidth]{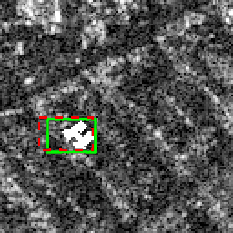}}}&
    \subfloat{{\includegraphics[width=.142\textwidth]{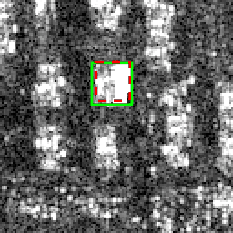}}}&
    \\\vspace{-0.4cm}\hspace{0.3cm}
    \rowname{\hspace{1.1cm}\scriptsize{RetinaNet$_h$}} 
    \subfloat{{\includegraphics[width=.142\textwidth]{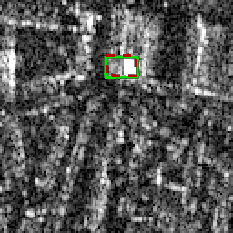}}}&
    \subfloat{{\includegraphics[width=.142\textwidth]{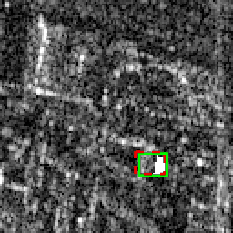}}}& 
    \subfloat{{\includegraphics[width=.142\textwidth]{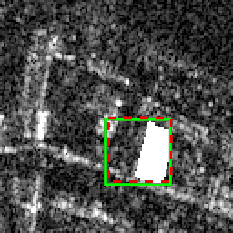}}}&
    \subfloat{{\includegraphics[width=.142\textwidth]{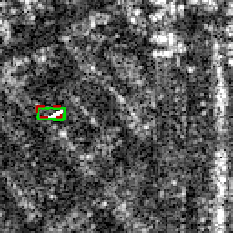}}}&
    \subfloat{{\includegraphics[width=.142\textwidth]{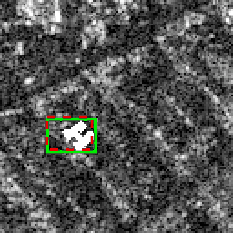}}}&
    \subfloat{{\includegraphics[width=.142\textwidth]{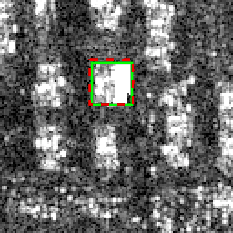}}}&
    \\\vspace{-0.4cm}\hspace{0.3cm}
    \rowname{\hspace{1.1cm}\scriptsize{Faster R-CNN$_h$}} 
    \subfloat{{\includegraphics[width=.142\textwidth]{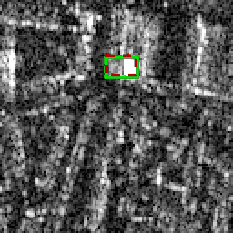}}}&
    \subfloat{{\includegraphics[width=.142\textwidth]{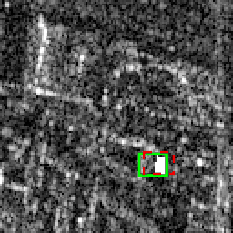}}}& 
    \subfloat{{\includegraphics[width=.142\textwidth]{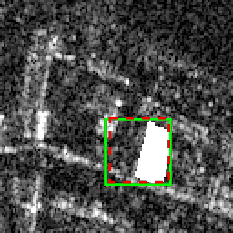}}}&
    \subfloat{{\includegraphics[width=.142\textwidth]{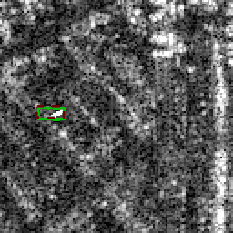}}}&
    \subfloat{{\includegraphics[width=.142\textwidth]{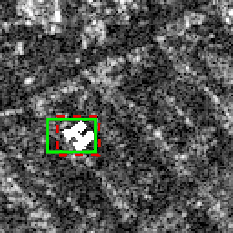}}}&
    \subfloat{{\includegraphics[width=.142\textwidth]{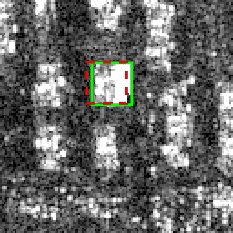}}}&
    \\\vspace{-0.4cm}\hspace{0.3cm}
    \rowname{\hspace{1.cm}\tiny{Faster R-CNN w.FPN$_h$}} 
    \subfloat{{\includegraphics[width=.142\textwidth]{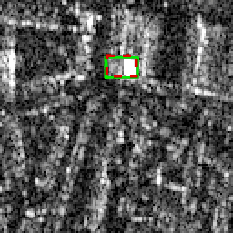}}}&
    \subfloat{{\includegraphics[width=.142\textwidth]{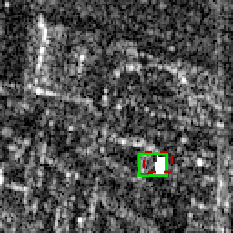}}}& 
    \subfloat{{\includegraphics[width=.142\textwidth]{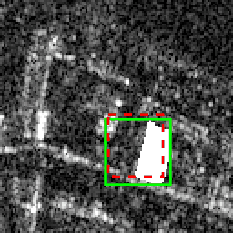}}}&
    \subfloat{{\includegraphics[width=.142\textwidth]{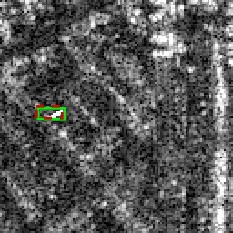}}}&
    \subfloat{{\includegraphics[width=.142\textwidth]{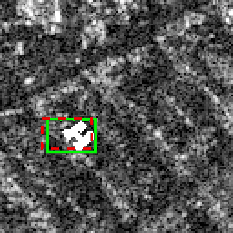}}}&
    \subfloat{{\includegraphics[width=.142\textwidth]{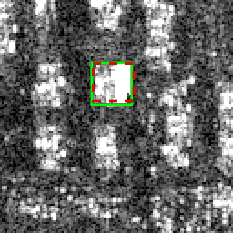}}}&
    \\\vspace{-0.4cm}\hspace{0.3cm}
    \rowname{\hspace{1.1cm}\scriptsize{Ours}} 
    \subfloat{{\includegraphics[width=.142\textwidth]{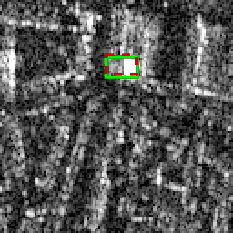}}}&
    \subfloat{{\includegraphics[width=.142\textwidth]{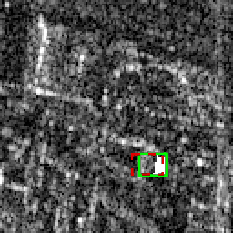}}}& 
    \subfloat{{\includegraphics[width=.142\textwidth]{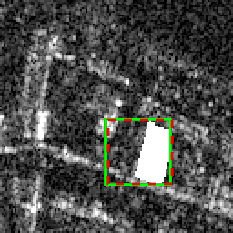}}}&
    \subfloat{{\includegraphics[width=.142\textwidth]{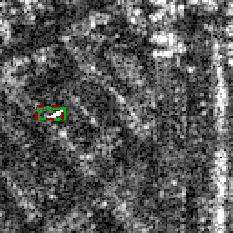}}}&
    \subfloat{{\includegraphics[width=.142\textwidth]{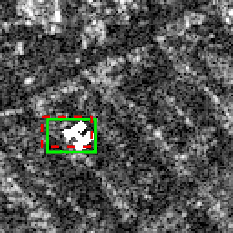}}}&
    \subfloat{{\includegraphics[width=.142\textwidth]{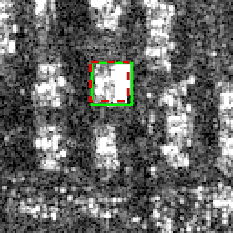}}}
    \vspace{0.22cm}
    \end{tabularx}
    \caption{Examples of predicted bounding boxes using different networks in Rotterdam and New York datasets. The predicted and ground truth bounding boxes are marked in red and green, respectively. 
    {The ground truth bounding boxes are obtained using the procedure described in Section 3.3. }
    }   \label{fig:sub_r_n}
\end{figure*} 

In addition to the quantitative evaluation, we visualize several predicted bounding boxes in Figure~\ref{fig:subgisF} and Figure~\ref{fig:sub_r_n}. 
In both the two figures, 
the first two rows show building footprint masks and SAR image patches, 
and Row 3 to 7 present predicted bounding boxes from each model, in which the corresponding building footprint masks and SAR images are both plotted. The ground truth boxes and predicted boxes are plotted in green and red, respectively. 

Figure~\ref{fig:subgisF} presents results of models in Berlin HS and Berlin SM data sets. 
We can observe a general improvement in quality from one-stage models to two-stage models and our network, especially for buildings in columns $b2$ and $b6$. 
All models can offer satisfactory results for buildings with larger footprints and clear signatures in the SAR image (e.g., the building in column $b4$). In contrast, for buildings with small footprints (see column $b1$) or ambiguous signatures (see column $b6$), one-stage models are not able to recognize full buildings. 
Besides, despite the resolution difference between the spotlight image and the stripmap image, the performance of all networks seems consistent. 

Figure~\ref{fig:sub_r_n} visualizes results of models in Rotterdam and New York data sets. 
Similar results can be seen in columns $b7$ and $b12$ that all networks perform well when building signatures clearly distinguish with surroundings. On the contrary, the predictions for the building in column $b8$ are not satisfactory. The same can be observed on column $b11$ in a building with a complex shape.  
Moreover, examples in columns $b9$ and $b10$ show two buildings are both well detected by all the networks despite the distinct differences in their footprints' sizes, 
which also indicates that FPN may not enhance the precision of predicted bounding boxes. 
In summary, the proposed network has a similar performance with Faster R-CNN$_h$.  

\begin{figure*}
    \subfloat{\includegraphics[width=\linewidth, clip=true,trim=4cm 0.65cm .5cm 4cm]{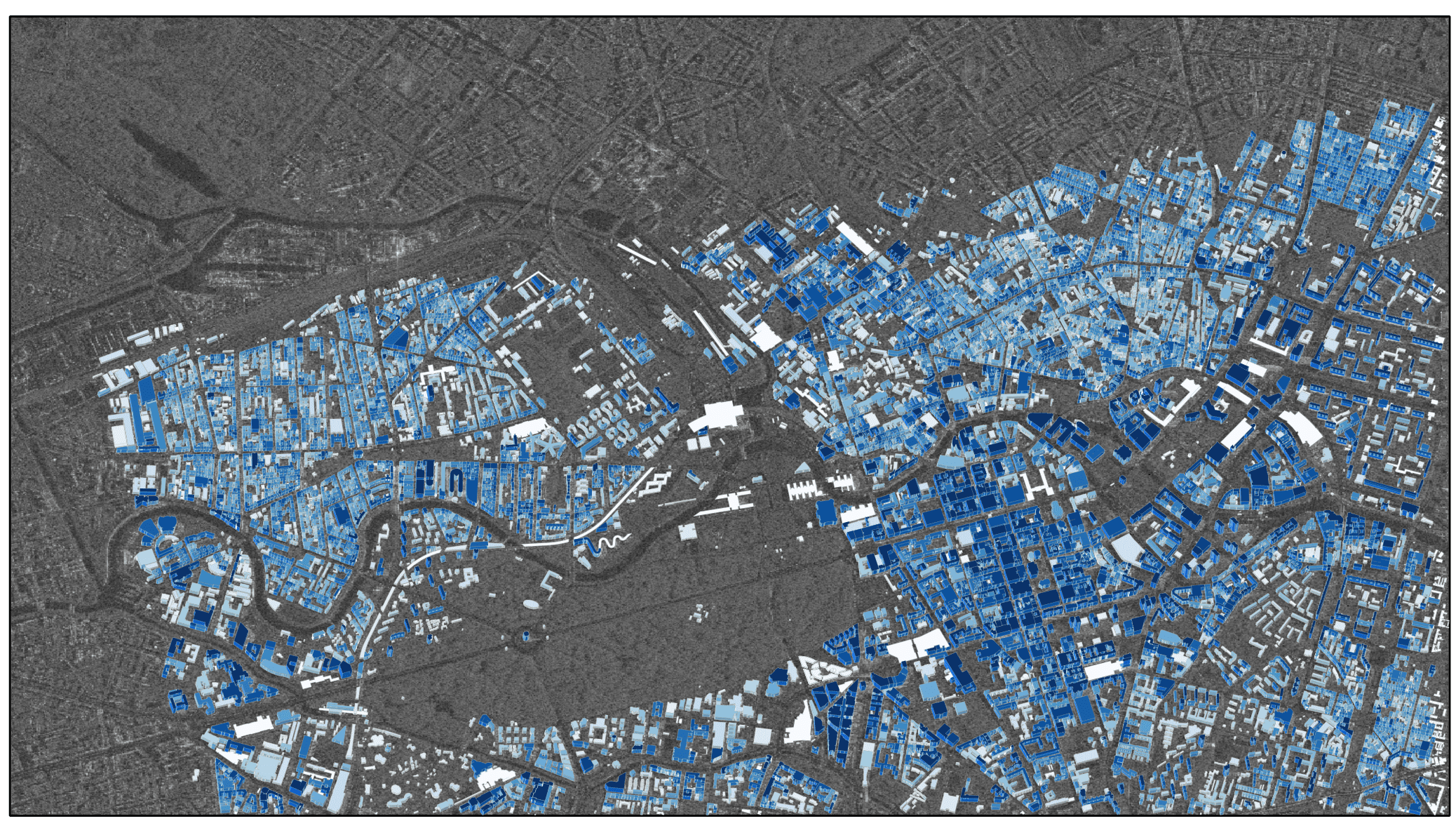}} \\
    \subfloat{\textcolor{gray}{\fboxrule=1pt\fboxsep=1pt\fbox{\includegraphics[width=.99\linewidth, clip=true,trim=4cm 0.65cm .5cm 4cm]{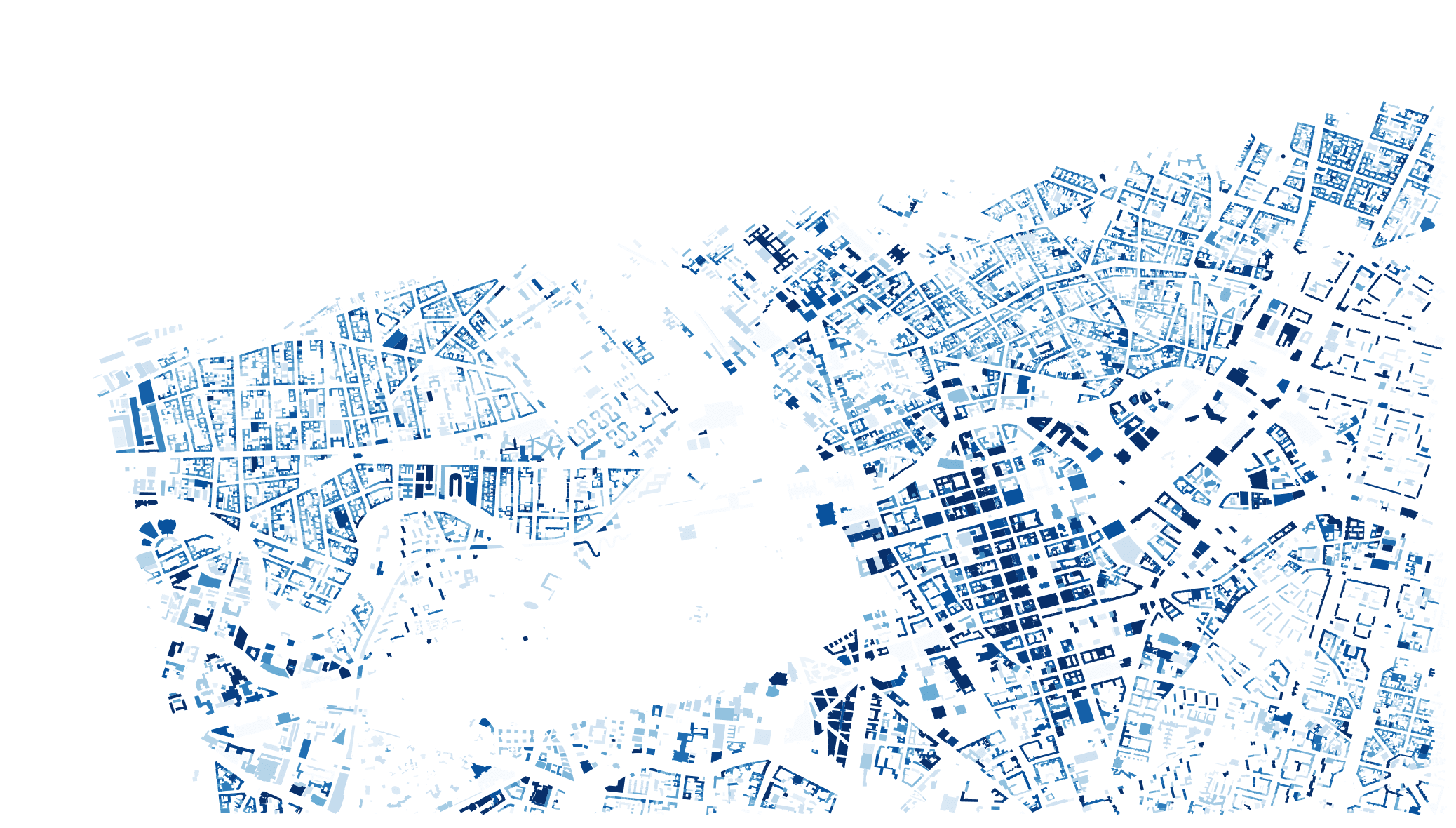}}}}
    \\
    \vspace{-0.2cm}
    \subfloat{\includegraphics[width=\linewidth]{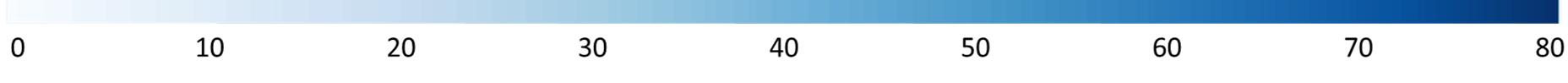}}
    \caption[Height prediction map in Berlin HS dataset.]
    {Height prediction map in Berlin HS dataset. 
    (up) Reconstructed LoD1 building models overlaid on the SAR image.  (down) Height prediction map in the SAR image coordinate system. Height is color-coded. }
    \label{fig:map_b_hs}
\end{figure*}

\begin{figure*}
    \subfloat{\includegraphics[width=\linewidth,height=0.4\linewidth,clip=true,trim=.8cm 0.2cm .8cm .7cm]{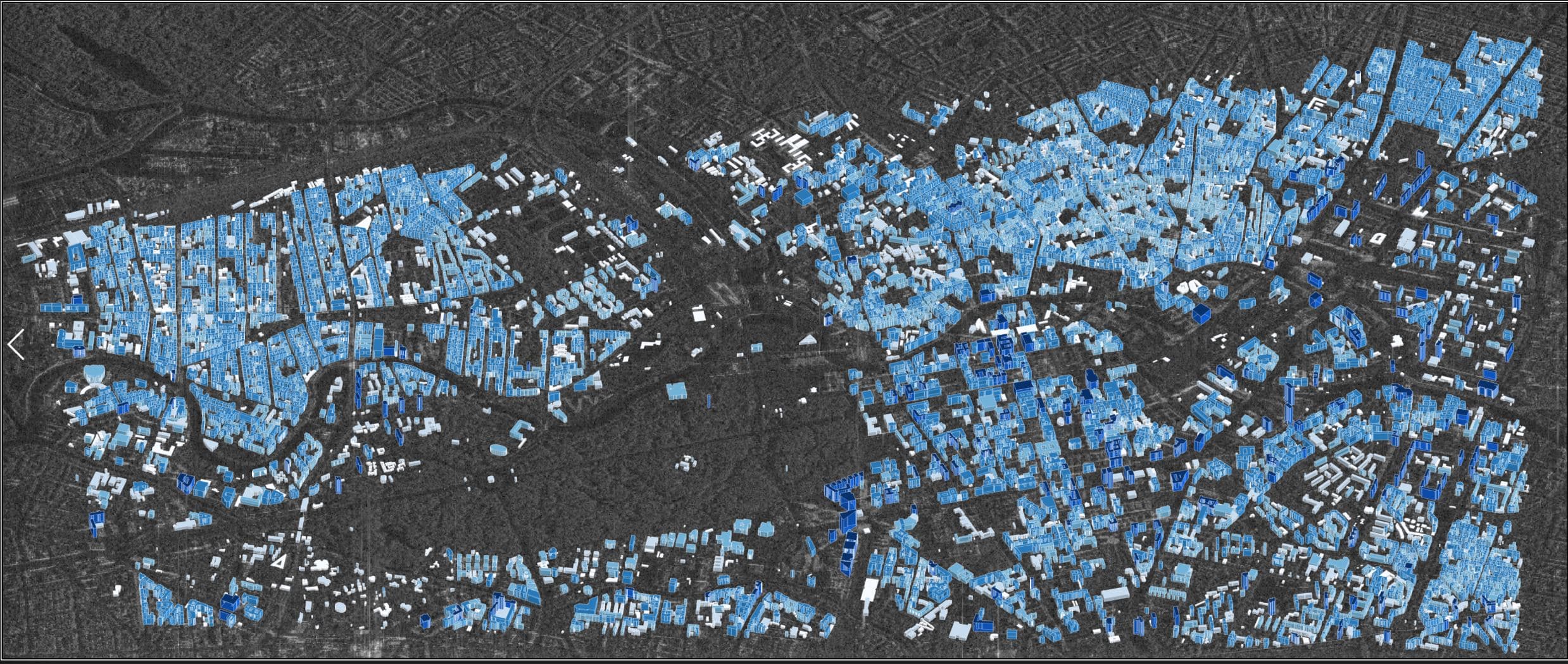}} \\
    \subfloat{\textcolor{gray}{\fboxrule=1pt\fboxsep=0pt\fbox{\includegraphics[
    width=\linewidth,height=0.4\linewidth]{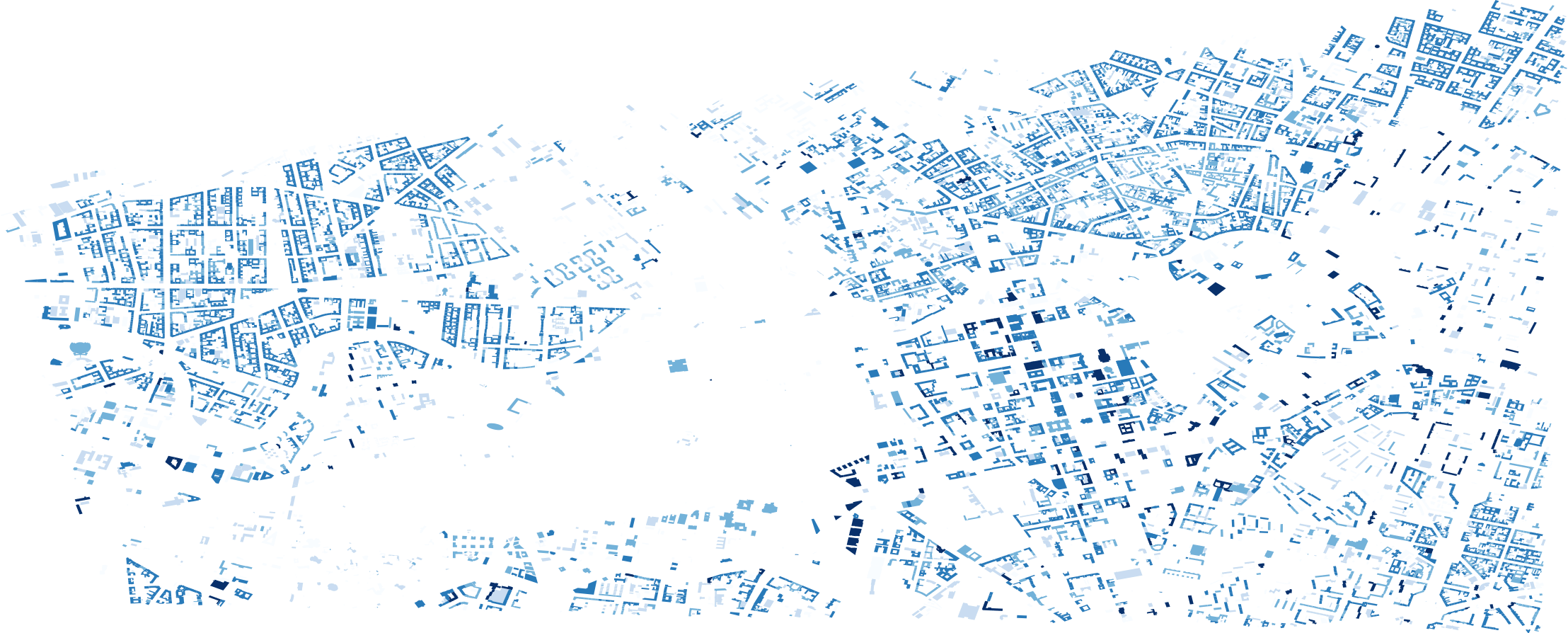}}}}
    \\
    \vspace{-0.2cm}
    \subfloat{\includegraphics[width=\linewidth]{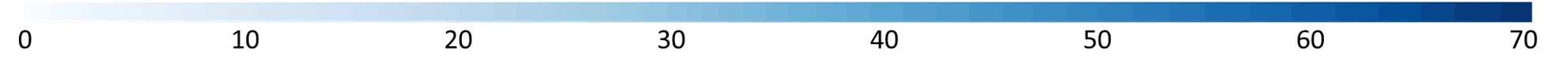}}
    \caption[Height prediction map in Berlin SM dataset. ]{Height prediction map in Berlin SM dataset. (up) Reconstructed LoD1 building models overlaid on the SAR image.  (down) Height prediction map in the SAR image coordinate system. Height is color-coded. }
    \label{fig:map_b_sm}
\end{figure*}

\begin{figure}
\pagestyle{empty}
    \centering
    \subfloat{\includegraphics[width=\linewidth,height=0.6\linewidth,clip=true,trim=0cm 0cm 1cm 0cm]{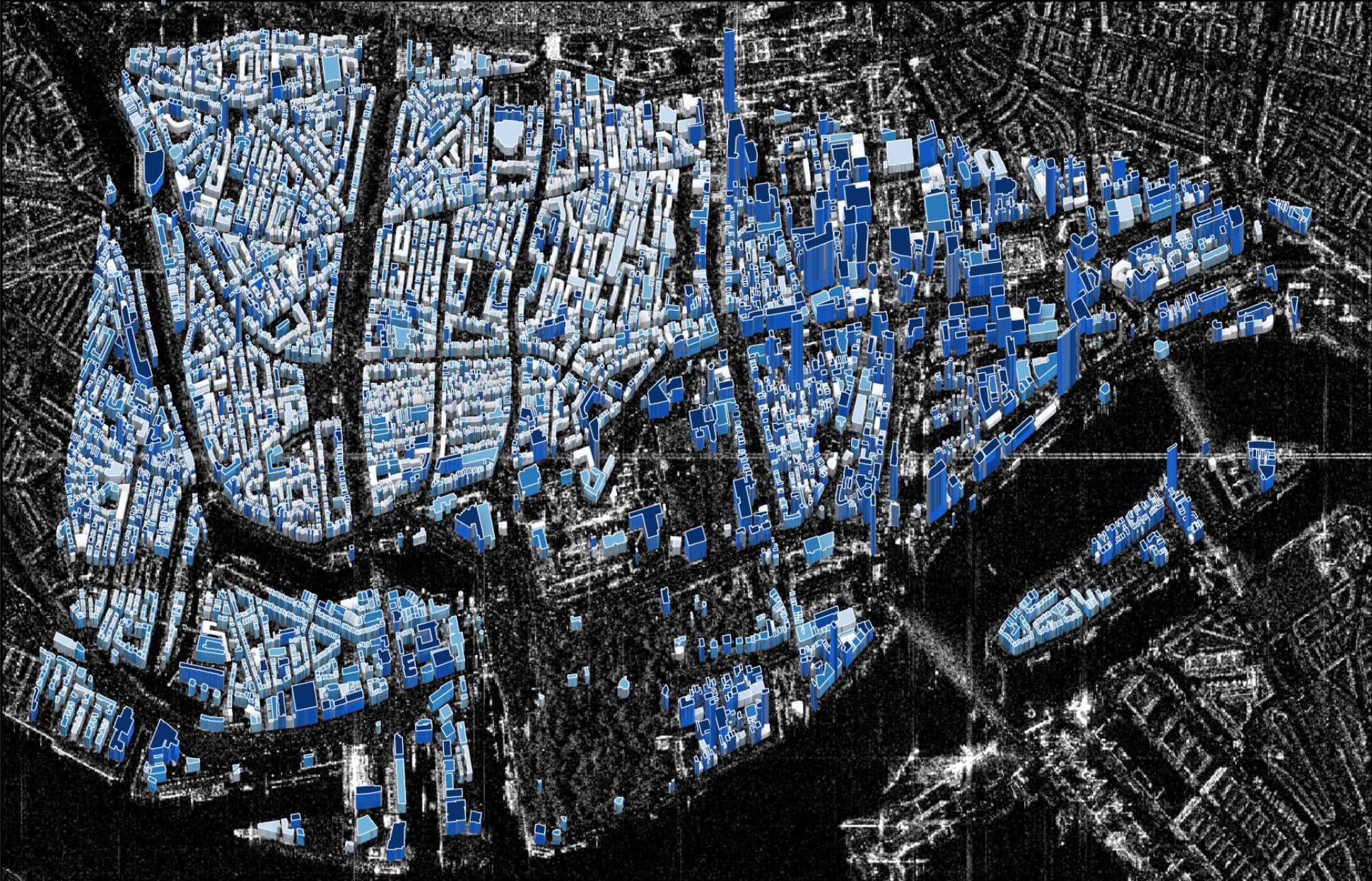}} 
    \\
    \subfloat{\textcolor{gray}{\fboxrule=1pt\fboxsep=9pt\fbox{
    \includegraphics[width=.951\linewidth,height=0.55\linewidth,clip=true,trim=0cm 0cm 0cm 0cm]{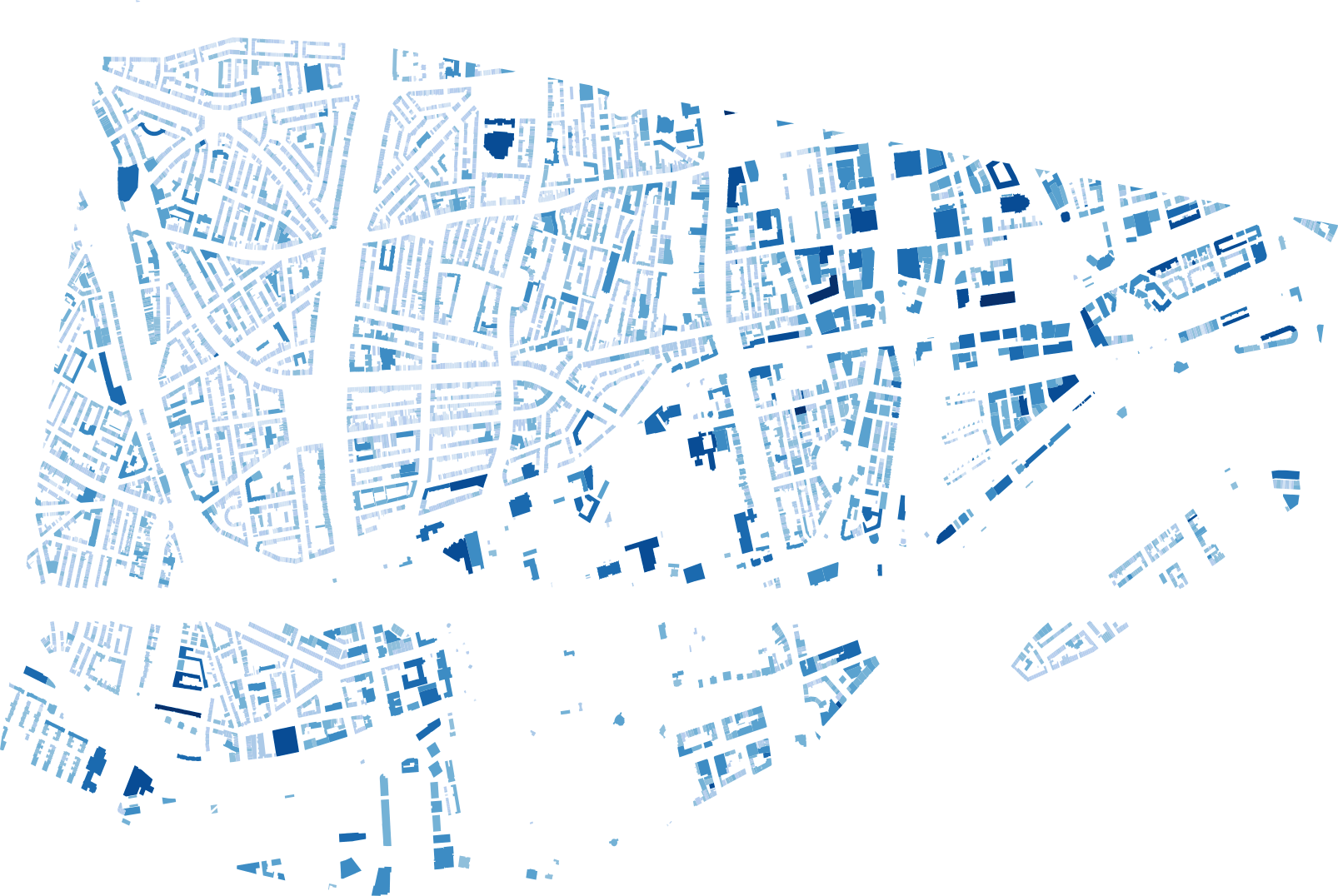}}}} %\\
    \\
    \vspace{-0.2cm}
    \subfloat{\includegraphics[width=\linewidth]{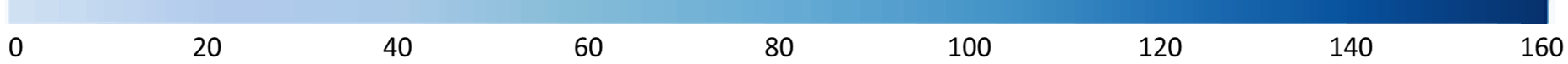}}
    \caption[Height prediction map in Rotterdam dataset. ]{Height prediction map in Rotterdam dataset.  (up) Reconstructed LoD1 building models overlaid on the SAR image.  (down) Height prediction map in the SAR image coordinate system. Height is color-coded. }
    \label{fig:map_rot}
\end{figure}

\begin{figure*}
    \centering
    \subfloat{\includegraphics[width=\linewidth,clip=true,trim=2cm 1cm 0cm 1cm]{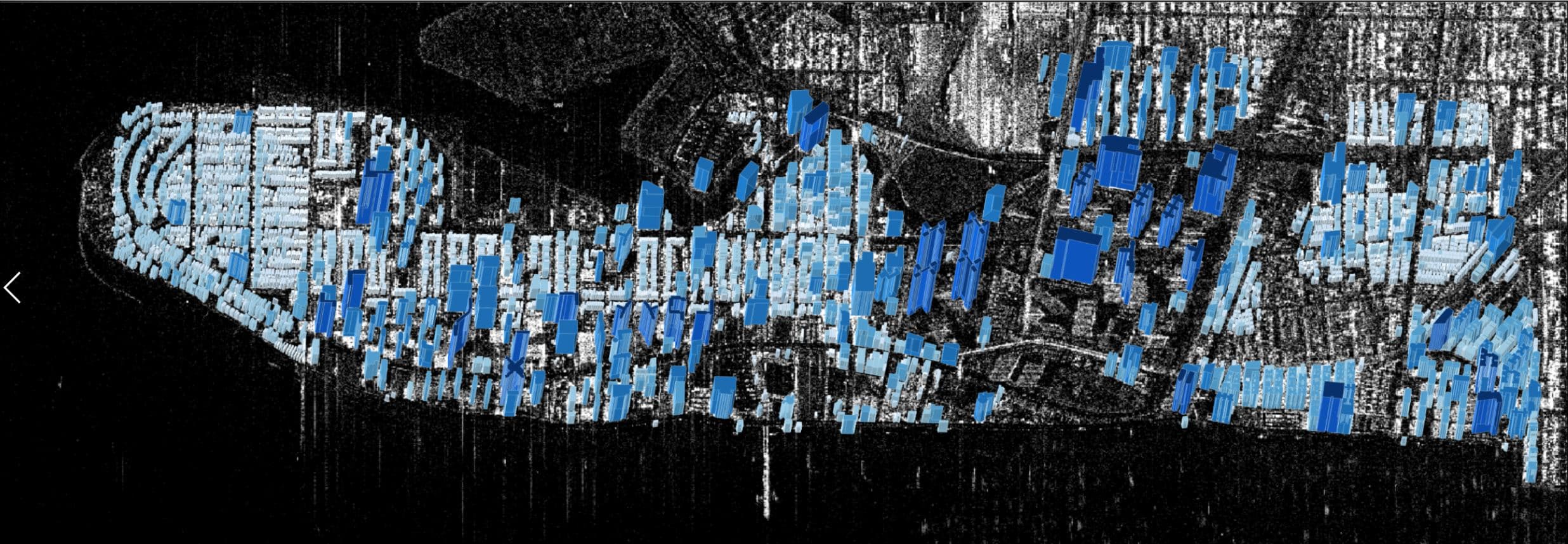}} \\
    \subfloat{%\colorbox{black}
    {\textcolor{gray}{\fboxrule=1pt\fboxsep=10pt\fbox{\includegraphics[width=.955\linewidth]{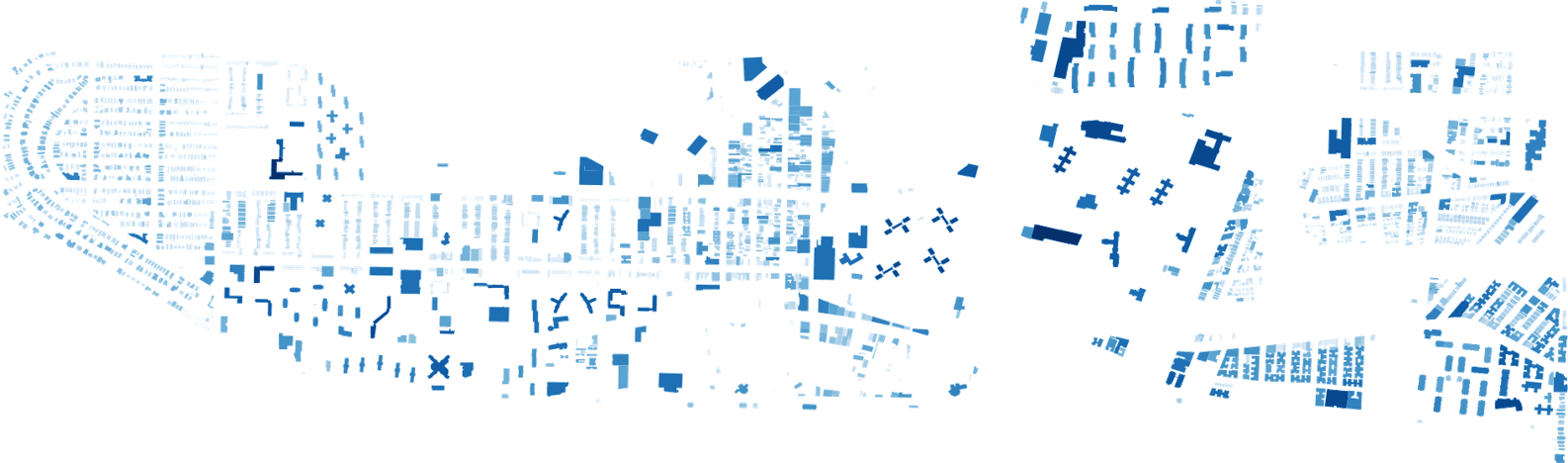}}}}}\\
    \vspace{-0.2cm}
    \subfloat{\includegraphics[width=\linewidth]{figs/ny_cb.png}}
    \caption[Height prediction map in New York dataset. ]{Height prediction map in New York dataset.  (up) Reconstructed LoD1 building models overlaid on the SAR image.  (down) Height prediction map in the SAR image coordinate system. Height is color-coded. }
    \label{fig:map_ny}
\end{figure*}

Reconstructed LoD1 building models and height prediction maps of our network in the whole areas of four data sets are plotted in %Figure~\ref{fig:map_b_hs}, Figure~\ref{fig:map_b_sm}, Figure~\ref{fig:map_rot},  Figure~\ref{fig:map_ny}. 
Figure~\ref{fig:map_b_hs} -- Figure~\ref{fig:map_ny}.

\section{Discussion}\label{sec:dis}

\subsection{Can our network work with inaccurate GIS data?}

So far, we have employed highly accurate building footprints in our experiments as they are acquired from official data sets. 
However, many openly available building footprints often contain positioning errors. To test the performance of the proposed network in such cases, we conduct supplementary experiments on training our proposed network with inaccurate building footprints and discuss the impact of positioning errors in GIS data.

According to a quality assessment study of OpenStreetMap (OSM) in~\cite{fan2014quality}, the average offset of building footprints is 4.13 m with a standard deviation of 1.71 m. 
We generate inaccurate footprints by injecting positioning errors to building footprints Ftp, resulting in inaccurate footprint data termed as Ftp-E. 
We choose Berlin HS data set for this experiment and term the data set containing positioning error in building footprints as Berlin HS-E. 

\begin{figure*}[!]
        \centering
        \includegraphics[width=.46\columnwidth]{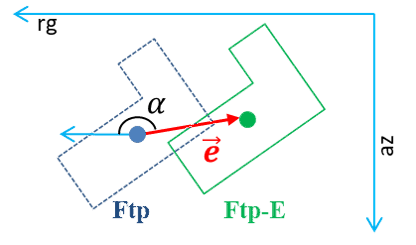}
        \caption{Illustration of generating building footprints with positioning errors. Positioning error {$\protect\overrightarrow{e}$} is added to building footprint Ftp, resulting in Ftp-E. rg and az denote the range direction and azimuth direction, respectively. $\alpha$ is the angle between {$\protect\overrightarrow{e}$} and rg. } 
        \label{fig:ftp_e} 
    \end{figure*}

The procedure is illustrated in Figure~\ref{fig:ftp_e}. 
$\protect\overrightarrow{e}$ is the added positioning error, and $\alpha$ is the angle between $\protect\overrightarrow{e}$ and the range direction. 
We consider the positioning error as a variable whose magnitude is Gaussian distributed, i.e., $|\protect\overrightarrow{e}| \sim \mathcal{N}(\mu=4.13, \sigma^2=1.71^2)$.   
Since the offset may point to different directions, we assume the direction of $\protect\overrightarrow{e}$ is uniformly distributed, i.e., $\alpha$  is uniformaly distributed in the range of $[0^{\circ},360^{\circ})$. 
For simplicity, let $\alpha$ be discrete: $\alpha \sim$ DiscreteUniform$(0^{\circ},359^{\circ})$. 
Note that this is the most difficult case that all footprints contain positioning errors. 
Then, we train our network on Berlin HS-E dataset and test the trained network with a clean test set. The parameter settings of the network remain the same as previous experiments, as described in Section~\ref{sec:train_details}.

\begin{figure*}[h!]
    \centering
    \settoheight{\tempdima}{\includegraphics[width=.15\columnwidth]{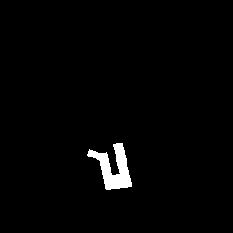}}%
    \begin{tabularx}{0.9\linewidth}{c@{\hskip3pt}c@{\hskip3pt}c@{\hskip3pt}c@{\hskip3pt}c@{\hskip3pt}}
    \\\vspace{-0.35cm}\hspace{1cm}
    \hspace{0.3cm}\footnotesize{{building a}} &{\footnotesize{{building b}}} 
    & {\footnotesize{{building c}}} & {\footnotesize{{building d}}} & {\footnotesize{{building e}}}
    \\\vspace{-0.4cm}\hspace{1cm}
    \rowname{\hspace{.1cm}\scriptsize{Footprint mask}} 
    \subfloat{{\includegraphics[width=.142\columnwidth]{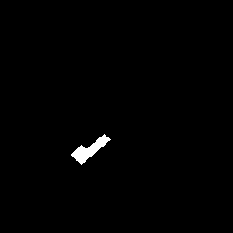}}}&
    \subfloat{{\includegraphics[width=.142\columnwidth]{figs/berlin_HS-E/02701_1_gis1.png}}}&
    \subfloat{{\includegraphics[width=.142\columnwidth]{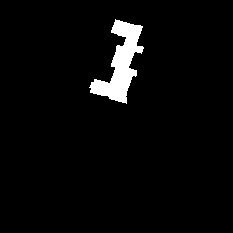}}}&
    \subfloat{{\includegraphics[width=.142\columnwidth]{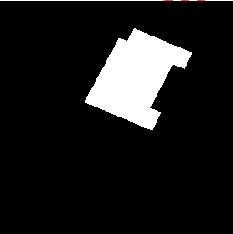}}}&
    \subfloat{{\includegraphics[width=.142\columnwidth]{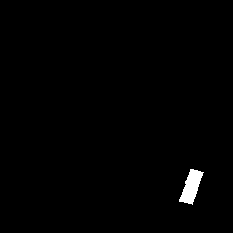}}}
    \\\vspace{-0.4cm}\hspace{1cm} 
    \rowname{\hspace{.1cm}\scriptsize{SAR image}} 
    \subfloat{{\includegraphics[width=.142\columnwidth]{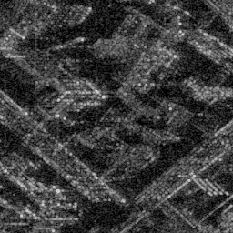}}}&
    \subfloat{{\includegraphics[width=.142\columnwidth]{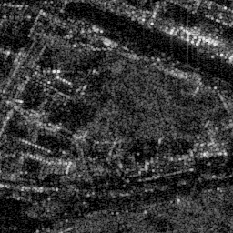}}}& 
    \subfloat{{\includegraphics[width=.142\columnwidth]{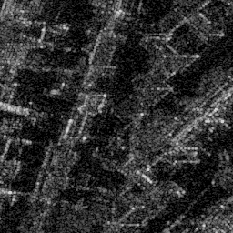}}}&
    \subfloat{{\includegraphics[width=.142\columnwidth]{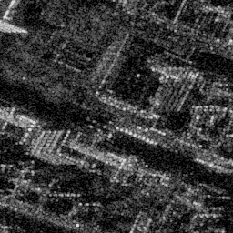}}}&
    \subfloat{{\includegraphics[width=.142\columnwidth]{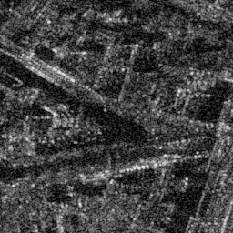}}}
    \\\vspace{-0.4cm}\hspace{1cm}
    \rowname{\hspace{.1cm}\scriptsize{Berlin HS}} 
    \subfloat{{\includegraphics[width=.142\columnwidth]{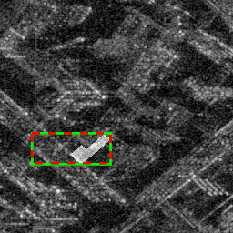}}}&
    \subfloat{{\includegraphics[width=.142\columnwidth]{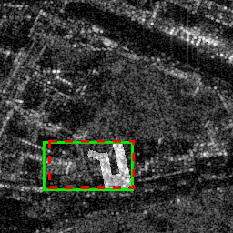}}}&
    \subfloat{{\includegraphics[width=.142\columnwidth]{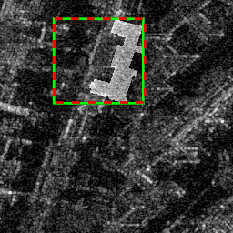}}}&
    \subfloat{{\includegraphics[width=.142\columnwidth]{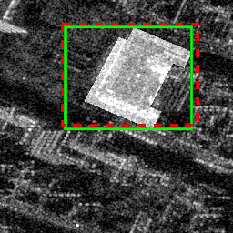}}}&
    \subfloat{{\includegraphics[width=.142\columnwidth]{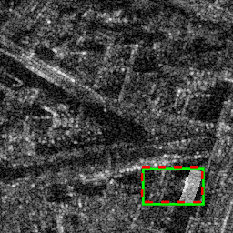}}}
    \\\vspace{-0.4cm}\hspace{1cm}
    \rowname{\hspace{.1cm}\scriptsize{Berlin HS-E}} 
    \subfloat{{\includegraphics[width=.142\columnwidth]{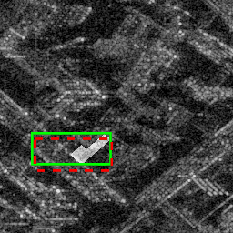}}}&
    \subfloat{{\includegraphics[width=.142\columnwidth]{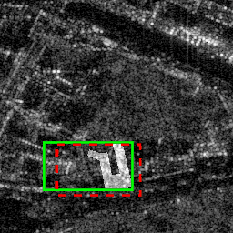}}}&
    \subfloat{{\includegraphics[width=.142\columnwidth]{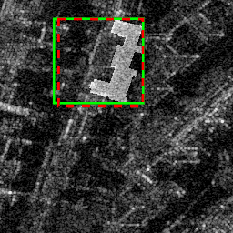}}}&
    \subfloat{{\includegraphics[width=.142\columnwidth]{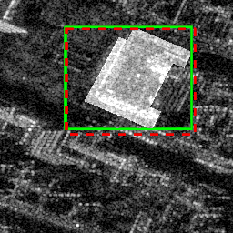}}}&
    \subfloat{{\includegraphics[width=.142\columnwidth]{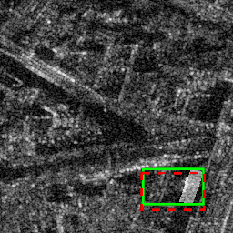}}}
    \\
    \end{tabularx}
    \vspace{0.4cm}
    \caption[Comparison of predicted bounding boxes using networks trained on building footprints with/without positioning errors. ]
    {Examples of predicted bounding boxes using networks trained with Berlin HS and Berlin HS-E (building footprints with positioning errors). 
    The predicted and ground truth bounding boxes and are marked in red and green, respectively. }
    \label{fig:berlin_hs_hse}
\end{figure*}

The results are listed in Table~\ref{tab:berlin_hse}.  
{As can be seen, compared to results obtained from Berlin HS, the mean height error is increased by 0.3 m, and the standard deviation of the height error is increased by 0.4 m.} However, it still gives competent height estimation results. 
For visual comparison, Figure~\ref{fig:berlin_hs_hse} shows the results of our network trained with Berlin HS and Berlin HS-E. 
As can be seen, for building b and c, the network trained with Berlin HS performs better. 
The predictions for buildings a and e are visually very similar. 
We observed that predictions from the network trained on Berlin HS-E are visually satisfactory for most buildings. 

The experiments show that the proposed network is robust against the positioning errors in building footprint data. 
This finding suggests that a large amount of existing open-sourced GIS data, such as OSM, can be exploited for height estimation of individual buildings in SAR images. 

% \begin{table}[!]
% \footnotesize
%     \centering
%     \caption{
%     Numerical results on trained using GIS and GIS-E on the Berlin HS data set.}\label{tab:berlin_hse}
%     \begin{tabular}{llll}
%     \hline
%     {Dataset} & {AP} & {{Height Error} (m)} &  {Training Time} \\
%                             &               & {mean} / {std}        &    \\
%     \hline
%     {\textit{Berlin HS}}   & {0.8917}       & 4.35 / 6.45           &    {1h58mins}\\
%     \hline
%     {\textit{Berlin HS-E}} & {{0.8031}}     & {4.77 / 6.88}         & {2h01mins} \\
%     \hline
%     \end{tabular}
% \end{table} 

\begin{table}[!]
\small
    \centering
    \caption{{
    Numerical results obtained from Berlin HS and Berlin HS-E data sets.}}
    \label{tab:berlin_hse}
    \begin{tabular}{lccc}
    \Xhline{2\arrayrulewidth}
    {Dataset}                &  $he_{mean}$ (m)  & $he_{std}$ (m) &   Training Time\\
    \hline
    {\textbf{Berlin HS}}        & 4.3 &  6.3      &      {1h01mins} \\
    \hline
    {\textbf{Berlin HS-E}}      & 4.6 &  6.8      &      {1h04mins} \\
    \Xhline{2\arrayrulewidth}
    \end{tabular}
\end{table} 

\subsection{{Influences of the nonlocal filtering procedure on SAR data}}

{In this work, we have employed original SAR amplitude images in our experiments. 
However, previous studies in ~\cite{shahzad2019Buildings, sun2020cgnet} perform nonlocal filtering~\cite{baierNonlocal} on SAR images prior to training to reduce the speckle effect. 
To test the influence of the nonlocal filtering procedure for our networks, 
we conduct supplementary experiments to train the proposed network with nonlocal filtered SAR images. }

{We perform denoising on SAR images using a nonlocal InSAR algorithm~\cite{baierNonlocal}. 
Berlin HS data set is chosen for this experiment, and the nonlocal filtered data set is termed Berlin HS-NL. 
Then, we train and test our network and all the comparative networks on Berlin HS-NL dataset. The parameter settings of the networks remain the same as previous experiments, as described in Section~\ref{sec:train_details} and Section~\ref{sec:comp_ex}.}

{Table~\ref{tab:nonlocal} lists the results. 
As can be seen, results from Berlin HS and Berlin HS-NL data sets are very similar on all networks. 
The experiments show that the filtering procedure does not improve the results. 
We think the reason might be that the large amount of filters in CNNs, in fact, have filtering effects on speckle noises. }

{This finding suggests that the filtering step is not needed for our task. Therefore, the computational cost for pre-processing can be largely reduced, which benefits especially for larger-scale processing. }

\begin{table*}[!]
\small
    \centering
    \caption{{
    Numerical results on four data sets. The highest values of different metrics are highlighted in {\textbf{bold}}. }
    }\label{tab:nonlocal}
    \begin{tabular}{llccc}
    \Xhline{2\arrayrulewidth}
    Data set &  Model Name       & $he_{mean}$ (m)  & $he_{std}$ (m) &   Training Time \\
    \hline
    \multirow{6}{*}{\textbf{Berlin HS}}   
        & SSD$_h$                        & 6.6 & 9.4   &          3h26mins \\ 
        &YOLOv3$_h$                     & 6.0 & 8.1   &           4h16mins \\
        &RetinaNet$_h$                   & 4.7 & 6.5   &           5h22mins \\
        &Faster R-CNN w.FPN$_h$          & 5.0 & 7.3   &           5h10mins\\
        &Faster R-CNN$_h$                & \textbf{4.3} & \textbf{6.2} & 5h26mins \\
        &Ours                         & \textbf{4.3} & 6.3   &        \textbf{1h01mins}\\
    \hline
    \multirow{6}{*}{\textbf{Berlin HS-NL}}   
        & SSD$_h$                        & 6.7 & 9.4   &          3h28mins \\ 
        &YOLOv3$_h$                     & 5.9 & 8.1   &          4h15mins \\
        &RetinaNet$_h$                   & 4.7 & 6.6   &           5h29mins \\
        &Faster R-CNN w.FPN$_h$          & 5.1 & 7.3   &           5h15mins\\
        &Faster R-CNN$_h$                & \textbf{4.3} & \textbf{6.4} & 5h28mins \\
        &Ours                         & \textbf{4.3} & 6.5   &        \textbf{1h04mins}\\
        \Xhline{2\arrayrulewidth}
    \end{tabular}
\end{table*}

\subsection{Pros and cons of retrieving building height using bounding box regression networks } 

We have applied the proposed network to four data sets and retrieved building heights.  
On Berlin HS data set, the mean height error is 4.3 m. 
In CG-net~\cite{sun2020cgnet}, 
the building height achieved using a segmentation network from the same SAR data is 2.39 m. 
The advantage of CG-net in terms of height accuracy is obvious. 
However, as aforementioned in Section~\ref{sec:intro}, 
pixel-wise labels are expensive, and it is not possible to generate training data for areas without accurate DEMs. Thus the applicability of CG-net is restricted.

The proposed regression network has two advantages. %are two folds. 
First, since the building height retrieval problem is formulated as a bounding box regression problem, the proposed method is capable of employing building height data from multiple sources. This enables the generation of annotation data on a larger scale and improves the transferability of the proposed networks. 
Second, the data set generation approach for bounding boxes is much simpler than the method for generating building areas in ~\cite{sun2020cgnet}. 
This is crucial when processing large data sets, e.g., on a regional or even larger scale. 

Comparing to the results of CG-net, building heights predicted using the proposed network have lower accuracy.   
The results are, however, still good. 
In~\cite{brunner2010Buildinga}, 
the authors retrieved heights of 40 isolated buildings from high-resolution spotlight TerraSAR-X images.  
For three building categories based on roof shapes, this work reported the mean height errors between -1 - 3.4 m and the standard deviation of height errors between 1.3 - 5.8 m. 
Our experiments employed both spotlight and stripmap SAR images and conducted performance testing on large amounts of data, for instance, 10K in Berlin HS data set (cf. Table~\ref{tab:trainTest}). 
Considering the image resolution of SAR data and the size of our data sets, 
the proposed method is very competitive. 

In summary, 
these comparisons suggest that the proposed bounding box regression network has great potential for applications aiming at large scales, e.g., to reconstruct baseline models on a regional or even global scale. When accurate DEMs are available, segmentation networks such as CG-net~\cite{sun2020cgnet} are preferred for higher accuracy on the reconstructed building heights. 

\section{Conclusion}\label{sec:conclude}

This work proposes a method that retrieves building heights in large-scale urban areas from a single TerraSAR-X spotlight or stripmap image. 
We formulate the problem of building height retrieval as a bounding box regression problem and develop a network that takes SAR images and building footprints as input and retrieves building heights by predicting building bounding boxes. 
Corresponding to the method, we propose a ground truth generation approach that only requires the footprint and one height value for each building. 
This approach can integrate multiple sources of building heights, such as open building models, LiDAR, and DEMs, thus can generate annotation data on larger scales and provide large potential in analyzing complex urban regions.

Four study sites are used to test the proposed networks, including one high-resolution spotlight TerraSAR-X image in Berlin and three stripmap TerraSAR-X images in Berlin, Rotterdam, and south Brooklyn in New York City. 
The mean height error achieved in the four sites ranges from 4.3 m to 5.7 m. 
The results are significant, as they are achieved from a single stripmap/spotlight TerraSAR-X image. 
Compared to methods utilizing object detection networks for building height retrieval, the proposed network can significantly reduce the computational cost while keeping the height accuracy of individual buildings compared to Faster R-CNN$_h$. 
Further experiments of training the networks using inaccurate building footprint data suggest that the proposed network is robust in the presence of positioning errors in building footprints, that a large amount of existing open-sourced GIS data, such as OSM, can be exploited for this task.

In the future, we are interested in improving the height accuracy of bounding box regression networks and producing LoD1 building models using SAR images in the stripmap mode on regional or even global scales. 
{We are also interested in domain adaptation techniques to mitigate the influences of the data differences between different areas and increase the transferability of our models.}

\section*{Acknowledgment}
This work is jointly supported by the European Research Council (ERC) under the European Union's Horizon 2020 research and innovation programme (grant agreement No. [ERC-2016-StG-714087], Acronym: \textit{So2Sat}), by the Helmholtz Association
through the Framework of Helmholtz AI (grant  number:  ZT-I-PF-5-01) - Local Unit ``Munich Unit @Aeronautics, Space and Transport (MASTr)'' and Helmholtz Excellent Professorship ``Data Science in Earth Observation - Big Data Fusion for Urban Research''(grant number: W2-W3-100) and by the German Federal Ministry of Education and Research (BMBF) in the framework of the international future AI lab "AI4EO -- Artificial Intelligence for Earth Observation: Reasoning, Uncertainties, Ethics and Beyond" (grant number: 01DD20001). 
The authors would like to thank Dr. H. Hirschmüller for providing the optical DEM.

%% `Elsevier LaTeX' style
\bibliographystyle{elsarticle-num}
\bibliography{mainbib}

\end{document}